\documentclass[twocolumn,prd,nofootinbib,aps,floats,floatfix,amsmath,amssymb,secnumarabic]{revtex4} %
\usepackage[final]{graphicx}
\usepackage{amsmath}
\usepackage{bbm}
\usepackage{amsfonts}
\usepackage{amssymb}
\usepackage{latexsym}
\usepackage{graphicx}
\usepackage[english]{babel}
\usepackage{multirow}
\usepackage{float}
\usepackage{url}
\usepackage{hyperref}
\usepackage{slashed}
\usepackage{xcolor} 

\def\sfrac#1#2{{\textstyle{#1\over #2}}}
\newcommand{\be}{\begin{equation}}
\newcommand{\ee}{\end{equation}}
\newcommand{\ba}{\begin{array}}
\newcommand{\ea}{\end{array}}
\newcommand{\bea}{\begin{eqnarray}}
\newcommand{\eea}{\end{eqnarray}}

\newcommand{\de}{{\mathbf e} }
\newcommand{\pd}{{\mathbf p} }
\newcommand{\dH}{{\mathbf H} }

\begin{document}

\title{Scattering properties of dark atoms and molecules}
\author{James M.\ Cline}
\author{Zuowei Liu}
\author{Guy D.\ Moore}
\affiliation{Department of Physics, McGill University,
3600 Rue University, Montr\'eal, Qu\'ebec, Canada H3A 2T8}
\author{Wei Xue}
\affiliation{INFN, Sezione di Trieste, SISSA, via Bonomea 265, 
34136 Trieste, Italy}

\begin{abstract} 
There has been renewed interest in the possibility that
dark matter exists in the form of atoms, analogous to those of the
visible world.  An important input for understanding the cosmological
consequences of dark atoms is their self-scattering.  Making use of
results from atomic physics for the potentials between hydrogen atoms,
we compute the low-energy elastic scattering cross sections for dark 
atoms.  We find an intricate dependence upon the ratio of the dark
proton to electron mass, allowing for the possibility to ``design''
low-energy features in the cross section.  Dependences upon other
parameters, namely the gauge coupling and reduced mass, scale out
of the problem by using atomic units.  We derive constraints on the
parameter space of dark atoms by demanding that their scattering cross
section does not exceed bounds from dark matter halo shapes.
We discuss the formation of
molecular dark hydrogen in the universe, and determine the analogous
constraints on the model when the dark matter is predominantly
in molecular form.
  
\end{abstract}
\maketitle

\section{Introduction}   Dark atoms are one of the oldest models of
particle dark matter \cite{Hodges:1993yb,Goldberg:1986nk},  originally
suggested by the venerable idea of mirror symmetry 
\cite{Berezhiani:1995yi}-\cite{Foot:2004pa}.  More recently, the idea
of dark sectors with gauge interactions not necessarily  identical to
those of the standard model has gained attention
\cite{Strassler:2006im,ArkaniHamed:2008qn}, motivating authors to 
take a fresh look at the implications of dark atoms 
\cite{Kaplan:2009de}-\cite{Cyr-Racine:2013fsa}.  Unlike  dark matter
consisting of elementary particles, dark atoms can have  large
self-interaction cross sections, which may impact   the structure of
galactic halo profiles, or those of clusters of galaxies, on which
there are observational constraints. 

A previous study
\cite{CyrRacine:2012fz} explored the impact of these constraints on
the parameter space of a simple atomic dark matter model, making
simplifying assumptions about the nature of the self-interaction cross
section.  In this paper we aim to avoid such assumptions and to
thereby obtain more accurate predictions, while illustrating a rich
range of possibilities for the energy-dependence of the cross
sections.   If tentative evidence for significant dark matter
self-interactions improves (for a recent review see
\cite{Weinberg:2013aya}), these features could prove useful
for model-building, as they allow one to construct scattering cross
sections with intricate features appearing at energies much lower than
would be possible in other theories of self-interacting dark matter.

We define  the atomic DM model in section \ref{model} and review
results from the atomic literature in the interaction potential
between atoms in section \ref{potentials}.  The methodology for
computing scattering cross sections is presented in section
\ref{formalism}, and the resulting predictions for scattering lengths
in the singlet and triplet channels are given in section
\ref{lengths}.  We present the energy-dependence of the atomic
cross sections in section \ref{xsections}.  Constraints on the model
from DM halo structure are derived in section \ref{constraints}.
In section \ref{analytic} we present analytic fits to the 
momentum-transfer cross section to facilitate the use of our results
by the reader.
The formation of dark molecules is discussed in section
\ref{molecules}, and the structure formation bounds analogous to those
of atoms are given in section \ref{mole_xsect}.  We summarize and
conclude in section \ref{conclusions}.

\section{The model}
\label{model} 

We assume that dark atoms ($\dH$) are analogous to visible hydrogen,
consisting of bound states of a fermionic dark electron $\de$ and
proton $\pd$ with masses $m_\de$ and $m_\pd$ respectively, and with
equal and opposite charges under a dark electromagnetism with massless
dark photon and fine structure constant $\alpha$.  (We do not refer to
the usual fine structure constant in this paper, so there will be no
confusion between the two.)  Otherwise the physics need not be the
same as in the visible sector, and in particular we do not assume that
there are dark neutrons or nuclei.  By definition, we take the dark
electron to be the lighter of the two constituents. The dark Hydrogen
binding energy is given by $\alpha^2\mu_\dH/2$ where  $\mu_\dH=m_\de
m_\pd/(m_\de+ m_\pd)$ is the reduced mass.  The atomic  unit 
(a.u.) of energy
is defined to be $\epsilon_0 = \alpha^2\mu_\dH$, while that of length
is the Bohr radius, $a_0 = (\alpha\mu_\dH)^{-1}$.  Sometimes we
will omit the explicit writing of $\epsilon_0$ and $a_0$ in the
specification of energies or distances; in such cases the use
of atomic units should be understood.

The model thus depends upon
only three parameters, which can be taken as $\epsilon_0$, $a_0$ and
the ratio $R = m_\pd/m_\de$.  We will see that the 
dependence of physical quantities on
$\epsilon_0$ and $a_0$ is trivial, if we ignore the  contribution of
the binding energy to the mass of the dark atom, $m_\dH \cong m_\de +
m_\pd$.  In this case they scale out of physical quantities by
choosing atomic units, leaving only $R$ as the relevant one to vary.

\section{Interatomic potentials}
\label{potentials}

\begin{figure}[t]
\includegraphics[width=\columnwidth]{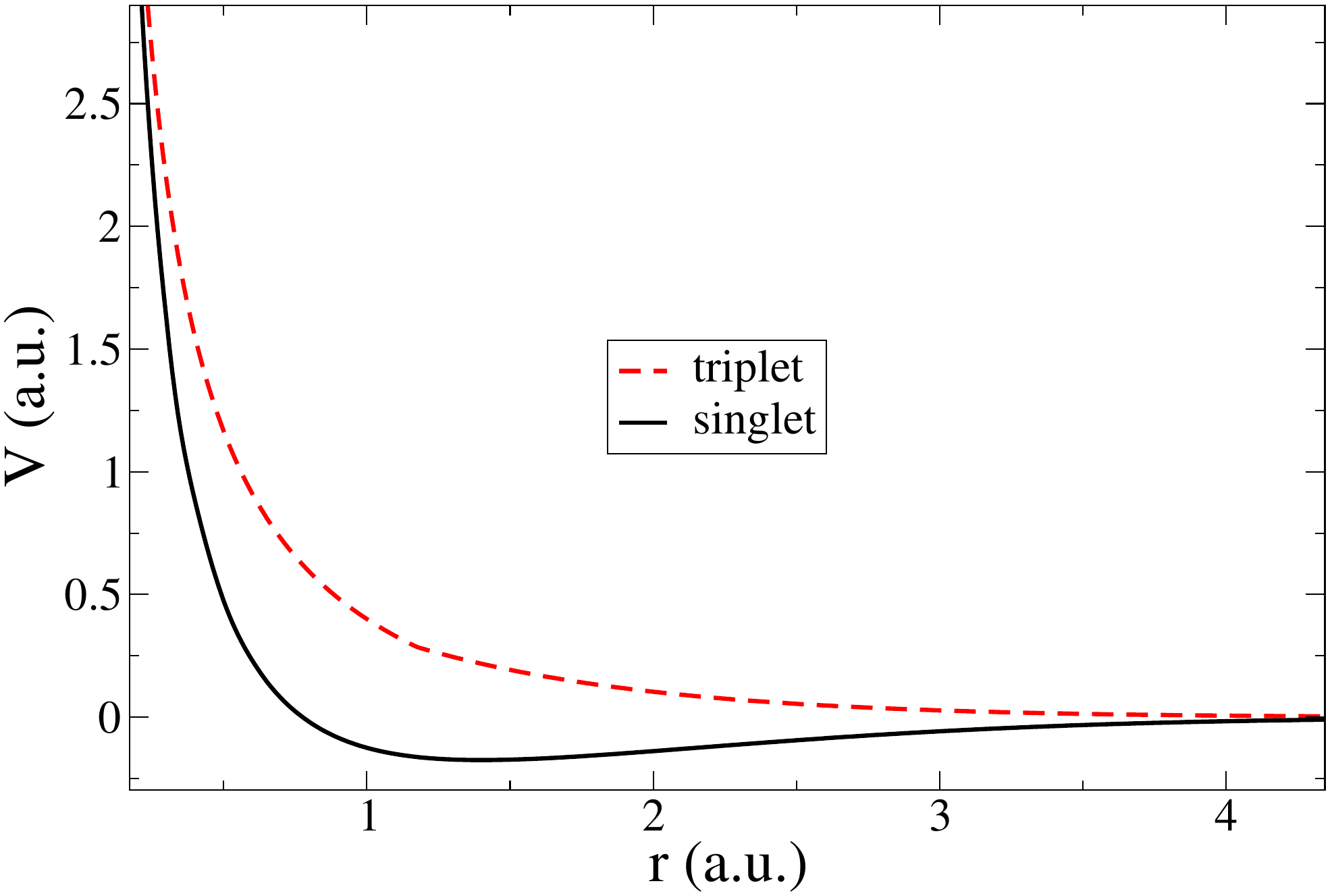}
\caption{Interaction potentials for hydrogen atoms with
electrons in spin singlet or triplet states.  Here and throughout,
``a.u.'' stands for ``atomic units,'' namely $\epsilon_0$ for energy
and $a_0$ for distance, as discussed in the text.
}
\label{fig:pot}
\end{figure}

The electrons in the scattering $\dH$ atoms can be in the spin
triplet or singlet states (which must be averaged over for unpolarized
scattering).  The interaction potential depends on the spin because the
overall wave-function must be antisymmetric.  The singlet state has a
symmetric spatial wave function, leading to a much deeper
potential well, allowing two $\dH$ to bind into molecular
$\dH_2$ with binding energy $0.16\, \epsilon_0$ and bond length
$1.4\, a_0$. (We are using the Born-Oppenheimer approximation, in which the
electronic state is solved for at each value of the $\pd$-$\pd$ separation,
or equivalently the $\de$'s respond adiabatically to the $\pd$ motion.
This approximation is valid in a large-$R$ expansion and we will make
it throughout.)  The triplet state must have a spatially 
antisymmetric electronic
wave function, leading to a potential with
a very shallow minimum with energy $-2\times 10^{-5}\,\epsilon_0$
at $r=7.9\,a_0$.  We plot them in fig.\ \ref{fig:pot}, and 
explain their origin in the following.

The triplet potential $V_t$ has been computed in \cite{KW74}, and an
analytic fit to these results has been given \cite{Silvera} as
\be
	{V_t\over\epsilon_0} = \exp(c_0 - c_1\, r - c_2\, r^2) 
       -D(r)\left({c_6\over r^6}+{c_8\over r^8}+ {c_{10}\over r^{10}}
	\right)
\label{silv}
\ee
with $c_0 = 0.09678$, $c_1 = 1.10173$, $c_2 = 0.03945$,
$c_6 = 6.5$, $c_8 = 124$, $c_{10} = 3285$, and 
$D(r) = \exp(-(r_1/r-1)^2)$ if $r<r_1$, $D(r)=1$ otherwise, with
$r_1=10.04$ in atomic units.
The first term represents the repulsive exchange
contribution, while the second models the attractive Van der Waals
part.  We find that (\ref{silv}) gives a good fit to the original
data of \cite{KW74} for $r>1.176$, and we extrapolate to lower $r$
using $V_t = -0.3652 + 0.7653/r$ for $r\le 1.176$.  (The $r^{-1}$
behavior provides a smooth fit to the tabulated potential at small
$r$.)

The singlet potential $V_s$ has been computed more recently in \cite{Wol93}
where results are tabulated in the range $0.2< r< 12$.  We interpolate
between the tabulated values for $0.3< r< 12$, and extrapolate to
small $r$ using $V_s =  -1.5379 + 0.94714/r$ for $r\le 0.3$.  To 
extrapolate to $r>12$, following \cite{Silvera} 
we have made a fit to $\ln(V_t-V_s)$ versus $r$, which turns out to
be nearly linear in $r$ in this region, thus obtaining
$V_s = V_t - \exp(2.3048 - 1.6238\,r)$.

We have found that the predictions for scattering from these
potentials are much more sensitive to small changes in $V_s$
than to $V_t$, due to the deeper minimum in the former.  It is 
therefore appropriate that more computational effort has been made 
in the atomic physics community to provide accurate recent
determinations of $V_s$, while the existing form of $V_t$ seems to be
adequate.  For example, ref.\ \cite{Joudeh} obtains a scattering
length from the approximation (\ref{silv}) that is consistent with
other studies.

\begin{figure*}[t]
\includegraphics[width=0.97\columnwidth]{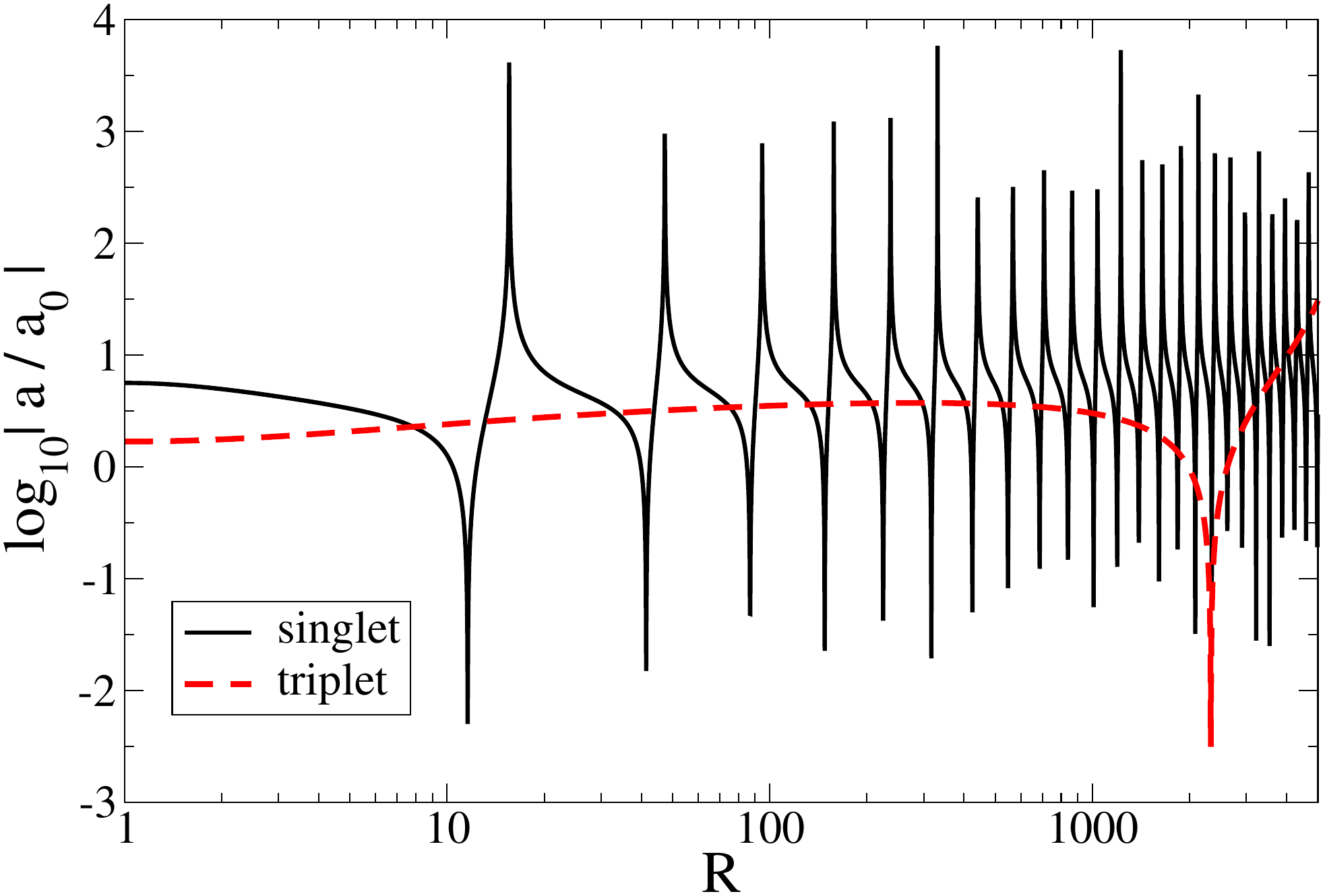}
\includegraphics[width=\columnwidth]{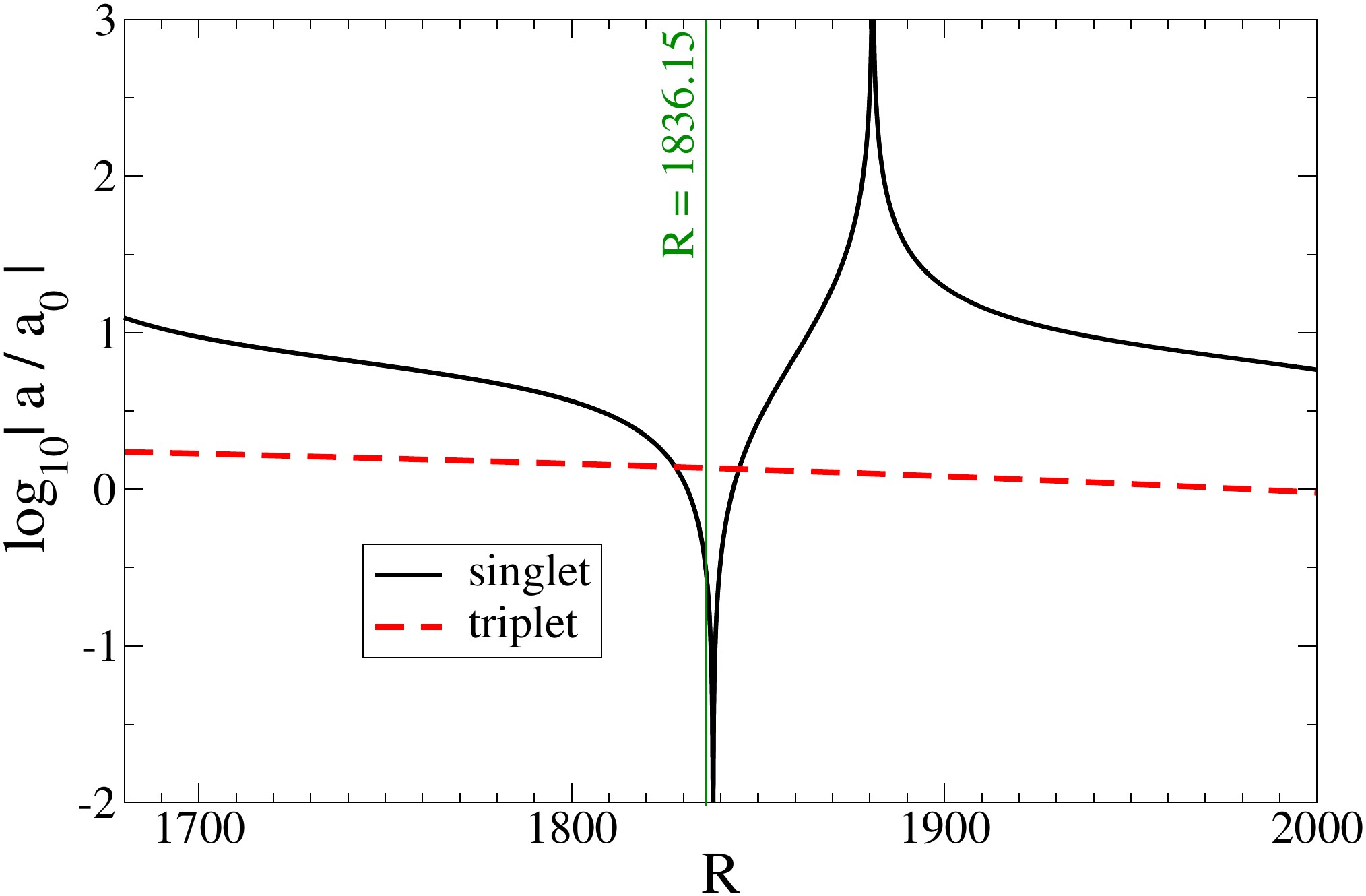}
\caption{Singlet and triplet scattering lengths as a function of
$R \equiv m_\pd/m_\de$.  
Rightmost figure zooms in on the region around $R=1836.15$, denoted
by vertical line.
}
\label{fig:slen}
\end{figure*}

\section{Scattering formalism}
\label{formalism}

To compute the elastic scattering properties of dark atoms, we
solve the Schr\"odinger equations for the partial wave radial
functions $u_\ell^{s,t} = r\,\psi_\ell^{s,t}$,
\be
	\left(\partial_r^2 -{\ell(\ell+1)\over r^2}
	+ f(R,\alpha)\,(E-V_{s,t})\right) u_\ell^{s,t}(r) = 0
\label{Schr}
\ee
where $\ell$ is the relative orbital angular momentum of the atoms, 
$r$, $E$ and $V_{s,t}$ are in atomic units, and 
\be
   f(R,\alpha) = m_\dH \, \epsilon_0\, a_0^2 = R + 2 + R^{-1}
	-\sfrac12\alpha^2
\label{feq}
\ee
is the ratio of the $m_\dH$ to $\mu_\dH$.
Here $E$ is the total c.m.\ energy of the colliding $\dH$ atoms,
and in the following we will ignore the binding energy ($\alpha^2$)
contribution to $f$ to approximate it as a function only of $R$.  It will be
useful to note that the wave number in atomic units is given by
$k = \sqrt{f E}$.

At distances large compared to the range of the potential,
$u_\ell^{s,t}$
takes the asymptotic form proportional to $\sin(kr - \ell\pi/2 
+\delta_\ell^{s,t}(k))$, where $\delta_\ell^{s,t}$ is the phase
shift.  The usual relation between the partial wave contribution to
the cross section and the phase shift is
$\sigma_\ell = (4\pi/k^2)(2\ell+1)\sin^2(\delta_\ell)$, 
but in the present case we must take into account the multiplicity
of the total nuclear spin, which is correlated to that of the 
electrons since the total wave function must be symmetric under
simultaneous interchange of both the electrons and the protons.  
Naively this would give extra relative weights of 
$(1/16,\,3/16)$ to the even- and odd-$\ell$ waves respectively
of the singlet state (since the nuclei must have total spin $0$
or $1$ respectively), while these weights would be $(9/16,\,3/16)$
for the triplet.  However it has been shown 
\cite{Jam99} that indistinguishability
of the two $\dH$ atoms gives rise to an additional factor of 2.
Then the expression for the total unpolarized cross section is
\be
	\sigma = {\pi\over 2\,k^2}\sum_{\ell}(2\ell+1)
\left\{\begin{array}{ll}  \phantom{9}\sin^2\delta^s_\ell + 
	 9\sin^2\delta^t_\ell,& \ell{\rm\ even}\\
	3\sin^2\delta^s_\ell + 3\sin^2\delta^t_\ell,&
	\ell{\rm\ odd}\end{array}\right.
\label{xsect}
\ee

To extract the phase shifts, one integrates the Schr\"odinger equation
from $r=\epsilon$ with $\epsilon \ll 1$ and $u_\ell(\epsilon)=0$
out to some sufficiently large $r$ where the $u_\ell$ is
well-approximated by the general $V=0$ solution, 
$u_\ell(r) =  C_1\, j_\ell(kr) + C_2\, n_\ell(kr)$.  These are
Ricatti-Bessel functions, related to the corresponding spherical
Bessel functions by a factor of $r$, so that their asymptotic behavior
is $j_\ell \sim \sin(kr-\pi\ell/2)$, $n_\ell \sim
-\cos(kr-\pi\ell/2)$.  At sufficiently large $r$, 
the coefficients are given by
\bea
   C_1 &=& -n'_\ell(kr)\,u_\ell(r) + n_\ell(kr)\,u'_\ell(r)/k
\nonumber\\
   C_2 &=& \phantom{-}j'_\ell(kr)\,u_\ell(r) - j_\ell(kr)\,u'_\ell(r)/k
\eea
where $k$ comes from the Wronskian, $j'_\ell = dj_\ell/d(kr)$, 
and the phase shift is then given by $\delta_\ell = \tan^{-1}(C_2/C_1)$.  One
can test for convergence by verifying that $\delta_\ell$ is
independent of $r$.  We find that $r=100$ is sufficient for energies
up to $E=0.1$ and $R<10^4$.

\section{Atomic scattering lengths}
\label{lengths}

In the limit $E\to 0$, the s-wave contribution to the
cross sections approach constant values
characterized by the scattering lengths
\be
	a_{s,t} = -\lim_{k\to 0} k^{-1}\tan\delta_0^{s,t}(k)
\ee
It can be calculated directly at $E=0$ by a simpler method than that
described for the phase shifts, since at $E=0$ the solution in the
region outside the potential is linear, $u_0 = C_2 - C_1 r$.  The
scattering length is the value of $r$ where this line intercepts
the $r$-axis:  $a = C_2/C_1 = \lim_{r\to\infty}[-u_0(r)/u_0'(r) + r]$. 
Again we find $r\sim 100$ adequate for our purposes. 

We first consider the proton-electron mass ratio $R=1836.15$ that
corresponds to the visible world.  We find scattering lengths
$a_s=0.28$ and $a_t=1.37$.  These are in reasonable agreement with
values found by other authors; for example, determinations of  $a_t$
in the atomic physics literature range from $1.2$ to $2$
\cite{Chakra06}.  Our values agree with those of \cite{Will95}. By
changing $R$ from 1836 to 1835, we can also reproduce the incorrect
value $a_s\cong 0.45$ obtained by several authors who neglected the
electron mass contribution to $m_\dH$ in (\ref{feq}), as has been
discussed in ref.\ \cite{Jam98}.

We next explore the dependence of the scattering lengths on the
proton to electron mass ratio $R$.  The results are shown in fig.\
\ref{fig:slen}.  The triplet scattering length $a_t$ varies 
relatively slowly with $R$, while the singlet one displays a large
number of poles and zeroes in the interval $R\in [1,5000]$.  This 
number is directly related to the number of bound states supported
by the corresponding potential.  
The dependence on $R$ can be
understood qualitatively from eq.\ (\ref{Schr}), which shows that the
potential effectively becomes deeper as $f(R)$ increases.  
A semiclassical analysis indicates that the number of bound states
should be of order $n \sim \pi^{-1}\int_{r_1}^\infty\sqrt{f(R)V}dr$ 
where $r_1$ is the $E=0$ turning
point; thus we expect that the $n$th pole of $a_s$ should occur at
$R_n \sim n^2$.  A numerical fit to the positions of the poles
confirms this, giving $R_n \cong -3.45 + 9.49\,n + 7.74\, n^2$.
The extreme
shallowness of the triplet potential is such that the first bound
state (first pole) only appears for $R > 2000$.  

Fig.\ \ref{fig:slen} shows a close-up of the region around $R\sim
1836$ corresponding to normal atoms.  It is a coincidence of nature
that we fall so close to a zero of $a_s$, so that the triplet
channel dominates even more than the $9:1$ ratio of coefficients
in (\ref{xsect}) would imply.  We will see below that as a result of
this accident, $R\sim 1836$ is close to a local minimum in the total 
cross section at low energy, considered as a function of $R$.

\section{Atomic cross sections}
\label{xsections}

\begin{figure}[b]
\includegraphics[width=\columnwidth]{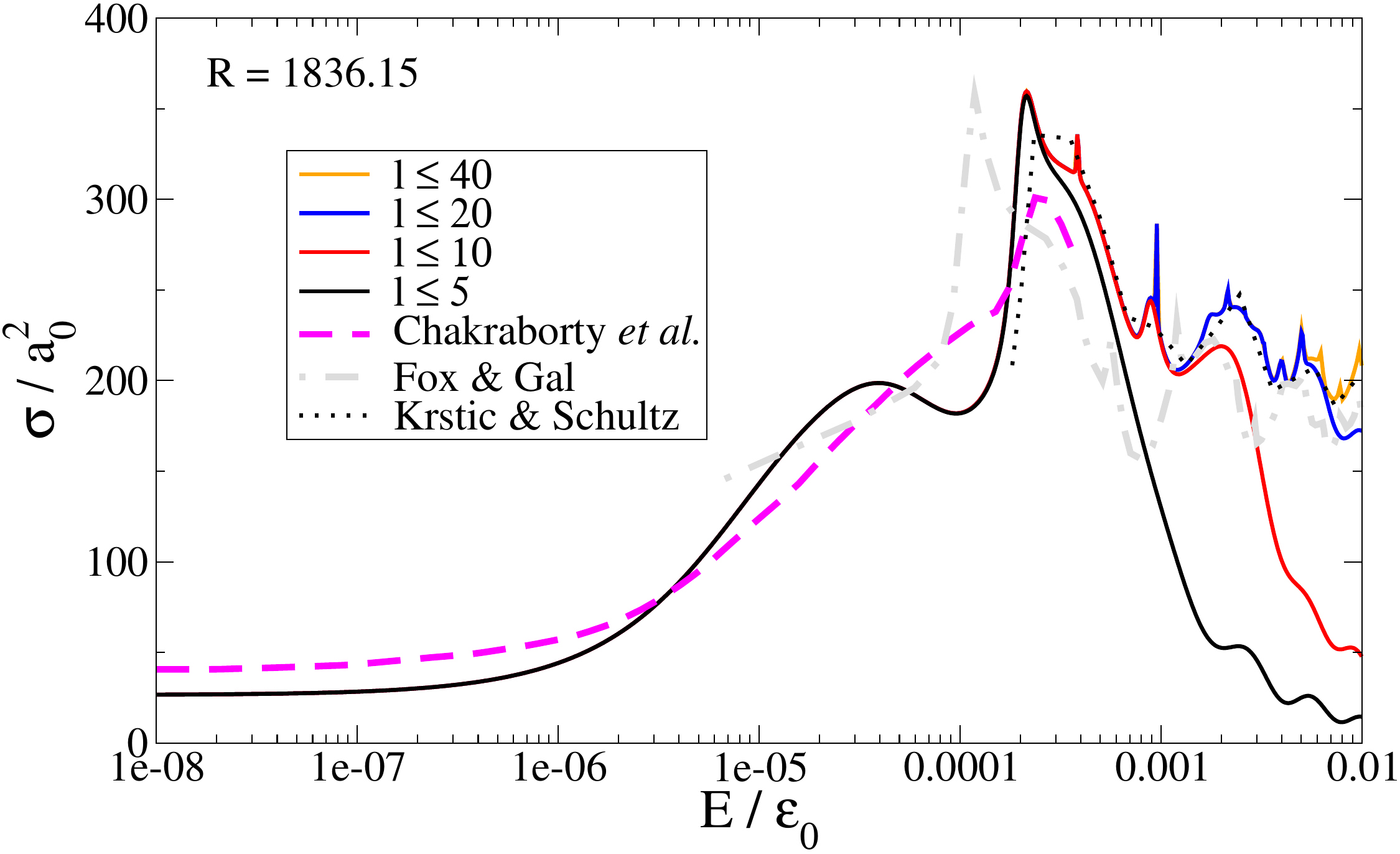}
\caption{Solid lines: our results for cross section with 
$R=1836.15$ including partial waves up to 
$\ell=5,\,10,\,20,\,40$.  Other curves: previous results from 
refs.\ \cite{Chakra07,Fox67,Krstic99}.
}
\label{fig:xsect1836}
\end{figure}

\begin{figure*}[t]
\includegraphics[width=2\columnwidth]{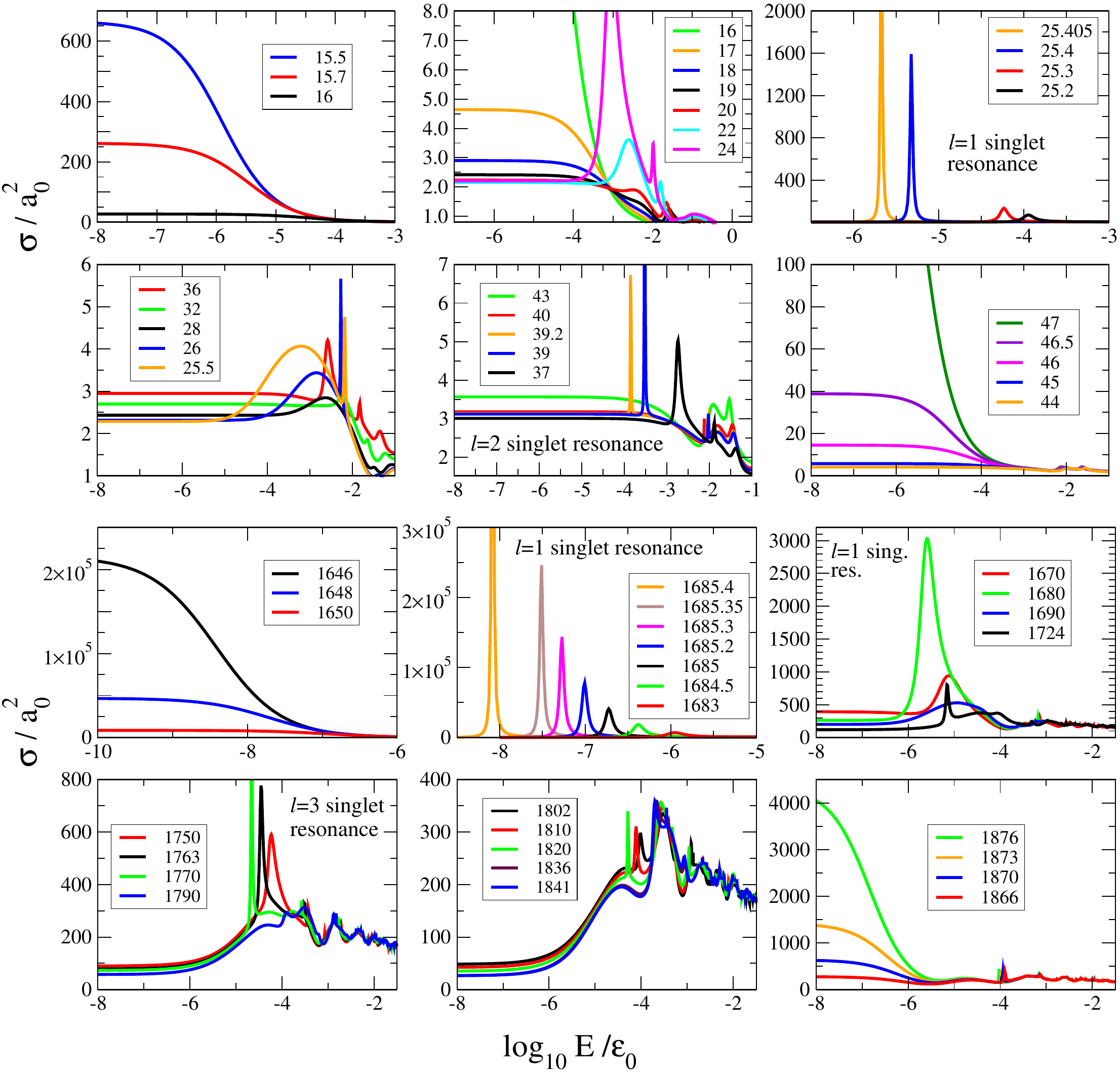}
\caption{Cross section as a function of energy for different values of
$R$, which are indicated in the legends.  For greyscale viewers, 
 the order of curves in the
legends corresponds to that of the low-energy parts of the curves,
from top to bottom.}
\label{fig:panels}
\end{figure*}

We now turn to the energy-dependent cross sections, exploring how they
change with $R$.  Our focus will be on low energies
$E\lesssim 10^{-2}\, \epsilon_0$, for which the cross sections converge
with the addition of a relatively small number of partial waves. At
higher energies, convergence can require including terms with $\ell$
in the hundreds.  For illustration and for comparison with previous
results in the literature, we start with the real-world value of
$R=1836.15$.  The result is shown in fig.\ \ref{fig:xsect1836} for
several different choices of the maximum partial wave number
$\ell_{\rm max}$.  For energies up to $E=10^{-3}$, $\ell_{\rm max}=10$
is sufficient.  For accurate predictions at $E=10^{-2}$, up to 40
partial waves are required at $R=1836$.  (At smaller $R$ we observe that
$\ell_{\rm max}=10$ is sufficient for all energies below $E=10^{-2}$.)

Fig.\ \ref{fig:xsect1836} also plots previous results for atomic H
scattering from references \cite{Chakra07,Fox67,Krstic99}. The results
are in reasonable agreement; in particular the  intricate structures
we obtain at energies $E > 2\times 10^{-4}$ match those of  ref.\
\cite{Krstic99} (dotted curve) very well.   The features can be
understood on general physical grounds. Unlike molecular bond lengths
and binding energies, which are expected to be determined by the
atomic units rather than $R$, the scattering cross section is
sensitive to the relation between energy and the deBroglie wavelength
of the atom, which involves the atomic, rather than electronic, mass,
and hence introduces $R$-dependence.  

Therefore, besides the energy scale $\epsilon_0$ that defines the
atomic unit of energy, there is a scale $\epsilon_1 = \epsilon_0
R^{-1}$ where the incoming atom's deBroglie wavelength is of order
$a_0$ and scattering becomes sensitive to the internal structure of
the atom, and another, $\epsilon_2 = \epsilon_0 R^{-3/2}$, where the
Van der Waals potential at a separation of one deBroglie wavelength is
of order of the kinetic energy.  It is the energy scale $\epsilon_2$
where scattering starts to change from being purely $s$-wave to
containing important contributions from higher partial waves.  In
fig.\ \ref{fig:xsect1836}, the cross section makes a transition from
flat to rising behavior at $E\sim 10^{-5}\sim \epsilon_2$, and it starts
falling again, while displaying numerous bumps and peaks,  around
$E\sim 10^{-3}\sim \epsilon_1$.

For general values of $R$, we find significant variations in the 
functional forms of $\sigma(E)$.  Fig.\ \ref{fig:panels} gives a
series of examples covering ranges  $R\in[15.5,\,47]$ and
$[1646,\,1876]$ that span neighboring poles of $a_s$ for
representative cases of lower and higher $R$. Although generically the
cross section approaches a constant as $E\to 0$, determined by the
scattering lengths, $\sigma \to (\pi/2)(a_s^2 + 9a_t^2)$, if $a_s$
diverges then $\sigma \sim 1/k^2$.  For values of $a_s$ close to a
pole, the transition between $\sigma \sim$ constant and $1/E$ 
behavior occurs at smaller energies than in the generic case.  This can
be seen in the graphs of fig.\ \ref{fig:panels} corresponding to
$R=15.5$, 47, 1646, and 1876.

A more generic behavior is illustrated by the plots corresponding to
$R=18,$ 36, 43, 1670, 1750, 1802; namely $\sigma$ remains close to its
asymptotic $E=0$ value until the scale $\epsilon_2$, and then starts
rising or falling, before entering the regime at $\epsilon_1$ where
rapid oscillations predominate, with a slowly falling envelope. 
Whether $\sigma$ falls or rises at $E=\epsilon_2$ depends on whether
$R$ is closer to being at a pole or a zero of $a_s$.  

Various resonances appear as $R$ is varied, but a particular one in
the $l=1$ singlet channel stands out, as is evident near $R=25$ and
$1685$.  It becomes more prominent and narrow as $R$ is increased up
to some critical value, at which point it abruptly disappears.  This
can be understood as the energy of the resonance passing through zero
at the critical value, after which it would only be seen for imaginary
values of the wave number, that of course we do not consider.  We
expect the resonance energies to decrease with $R$ since increasing
$R$ makes the potential deeper (see eq.\ (\ref{Schr})), causing all the energy levels to go
down.  This is also true for the positive-energy virtual states, which
become negative-energy bound states as $R$ increases.

\begin{figure}[t]
\includegraphics[width=\columnwidth]{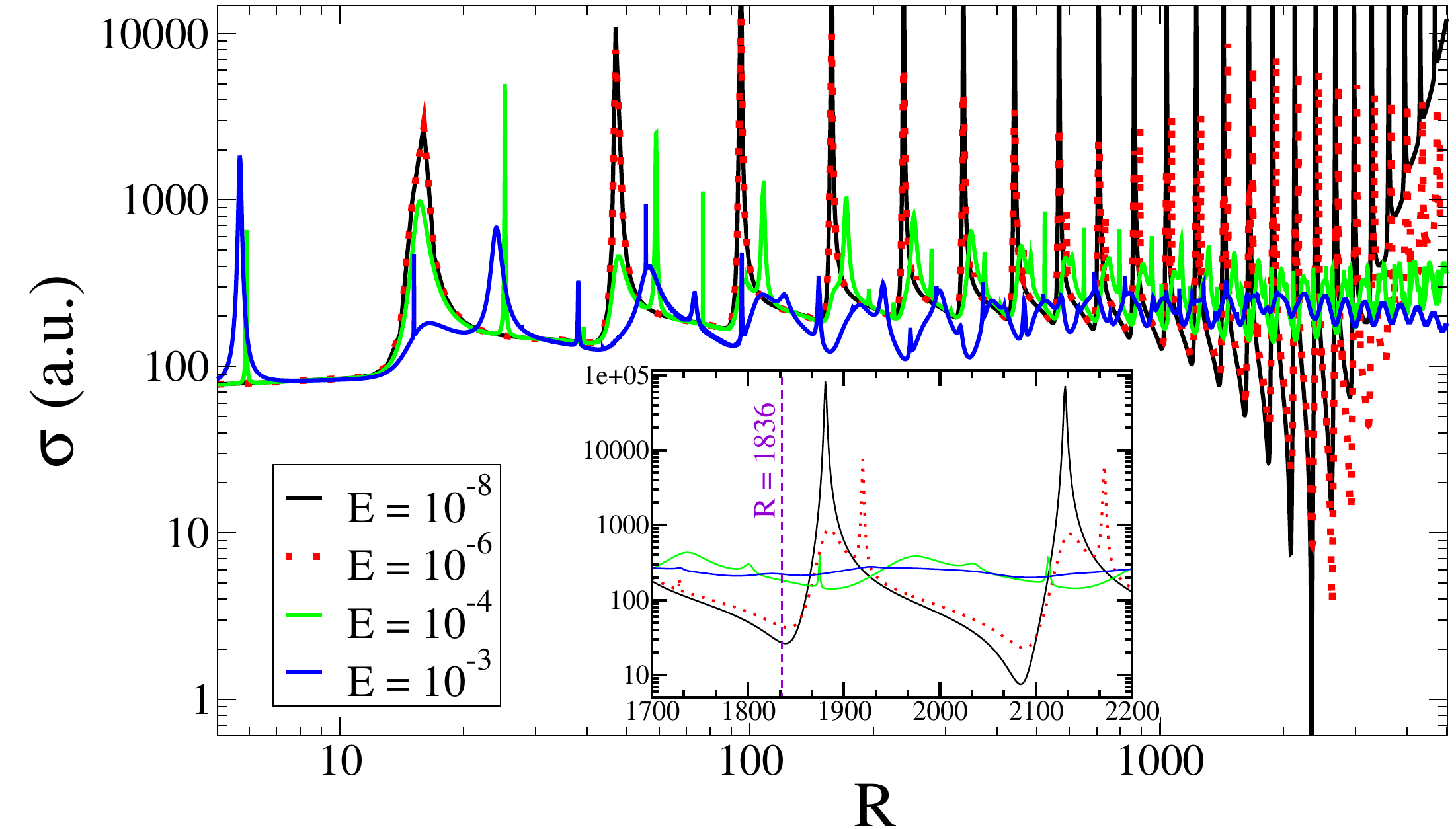}
\caption{Cross section as a function of $R$ for 
different fixed values of the energy, 
$E=10^{-3}$, $10^{-4}$, $10^{-6}$ and
$10^{-8}$ in atomic units. (Lower energies generally correspond to
larger $\sigma$.)  Inset zooms in on the region near $R=1836$.}
\label{fig:low-E}
\end{figure}

The global behavior of $\sigma(E)$ as a function of $R$ can also
be visualized by plotting $\sigma$ versus $R$ at a few fixed energies.
We show this for energies $E=10^{-3}$, $10^{-4}$, $10^{-6}$ and
$10^{-8}\, \epsilon_0$ in fig.\ \ref{fig:low-E}.  Generally we
observe a minimum cross section of order $\sigma\sim 100\, a_0^2$, 
except in the region $R\sim 2000-3000$ near the first zero of the
triplet scattering length.  Curiously the natural value $R=1836$
is at a local minimum of the total cross section, as can be
seen in the inset of the figure.  Only three other zeroes of $a_s$
correspond to such a low value of $\sigma$.

\section{Dark atom constraints from galactic structure}
\label{constraints}
Self-interactions of dark matter have been studied in connection with
their effects on structure formation within galaxies and galactic
clusters, leading to upper bounds on the cross section.  The
constraints come about because dark matter  tends to be
slower in the periphery of a bound structure, and the interactions
between particles with large- and small-radius orbits can heat up the
interior particles, causing them to escape to wider regions, and 
leading to cored profiles for galaxies.   On
larger scales, the observed ellipticity of halos in clusters will be
erased by strong self-scattering, leading to spherical halos. 

Formerly, halo ellipticity was believed to give the strongest bound,
$\sigma/m < 0.02\,$cm$^2/$g \cite{MiraldaEscude:2000qt} for DM with
velocities of order 1000 km/s, characteristic of galactic clusters. 
But recent studies based upon $N$-body simulations 
\cite{Peter:2012jh} have concluded that the true bound is much weaker,
at least as large as $0.1\,$cm$^2/$g, but smaller than $1\,$cm$^2/$g
\cite{Peter:2012jh}.  The latter is consistent with similar bounds
obtained from the Bullet Cluster
\cite{Markevitch:2003at,Randall:2007ph} and from accretion of dark
matter by supermassive black holes in galactic centers 
\cite{Ostriker:1999ee}. (Ref.\ \cite{Hennawi:2001be}  obtains a
stronger  constraint, which however depends upon assuming a cuspy
profile for the DM halo, which might be erased by the
self-interactions themselves.)  Ref.\ \cite{Gnedin:2000ea} constrains
$\sigma/m < 0.4\,$cm$^2/$g from requiring that elliptical galaxy halos
within clusters do not evaporate within $10^{10}\,$y, at 
DM velocities of $v\sim 100\,$-$1000\,$km/s, 
while ref.\ 
\cite{SanchezSalcedo:2005vc} obtains $\sigma/m < 0.2\,$cm$^2/$g from
the inferred DM profile of a particular low surface-brightness galaxy
with $v\sim 150\,$km/s,
with input from then-current cosmological simulations.  More recently
it has been argued \cite{Zavala:2012us} that a value of $\sigma\sim
0.6\,$cm$^2/$g would be consistent with observed central densities of
the Milky Way dwarf spheroidals at $v\sim 10\,$km/s.  Taking into
account the probable though unspecified astrophysical uncertainties in these bounds, a 
reasonable and simple compromise would seem to be $\sigma/m < 
0.5\,$cm$^2/$g \cite{Tulin:2013teo}, 
which we adopt in the following.  We will apply this bound over the
range of velocity scales $v \in [10,1000]$ km/s that are relevant
for dwarf spheroidals up to galactic clusters in the following.

Scatterings in the forward direction are not
effective for exchanging energy between dark matter particles, which
is the basis for the constraints on $\sigma$.  Therefore the 
bound should be applied not to the elastic cross
section $\sigma_{\rm el}$, but rather to the transport cross section $\sigma_t$, which
gives a better representation of scatterings that involve significant
exchange of momentum.   A commonly used expression for the transport
cross-section is
$\sigma_{t} = 2\pi\int d(\cos\theta) (1-\cos\theta)d\sigma/d\Omega$.
But this expression is not appropriate for scattering between identical
particles, since backward scattering is indistinguishable from forward
scattering and is also not effective at modifying the momentum distribution.
Therefore it is more appropriate to use
$\sigma'_{t} = 2\pi\int d(\cos\theta) (1-\cos^2\theta)d\sigma/d\Omega$,
which treats forward and backward scattering as equivalent.%
\footnote{%
    Here we disagree with refs.\ \cite{CyrRacine:2012fz}
    and \cite{Krstic99}, which use $\sigma_t$ and therefore find that the
    elastic and transport cross sections are equal for identical particles.
    Their approach is based on treating forward scattering as irrelevant
    ($1-\cos(\theta)=0$) but backwards scattering as of maximal relevance
    ($1-\cos(\theta)=2$), which is inconsistent since for identical particles
    these processes are equivalent.
    }
In terms of
partial waves, it is given by \cite{PDR03}
\be
	\sigma'_{t} = {6\pi\over k^2}\sum_\ell
{(\ell+1)(\ell+2)\over(2\ell+3)}\sin^2(\delta_{\ell} - 
\delta_{\ell + 2})
\ee
for a generic scattering problem.  For the current application,
eq.\ (\ref{xsect}) is adapted by replacing $(2\ell+1)$ with
$(3/2)(\ell+1)(\ell+2)/(2\ell+3)$ and $\delta_{\ell}$ by 
$\delta_{\ell} - \delta_{\ell + 2}$.  The normalization is such that
at low energies where only the $s$-wave contributes, $\sigma'_t =
\sigma_{\rm el}$.  

\begin{figure}[t]
\includegraphics[width=\columnwidth]{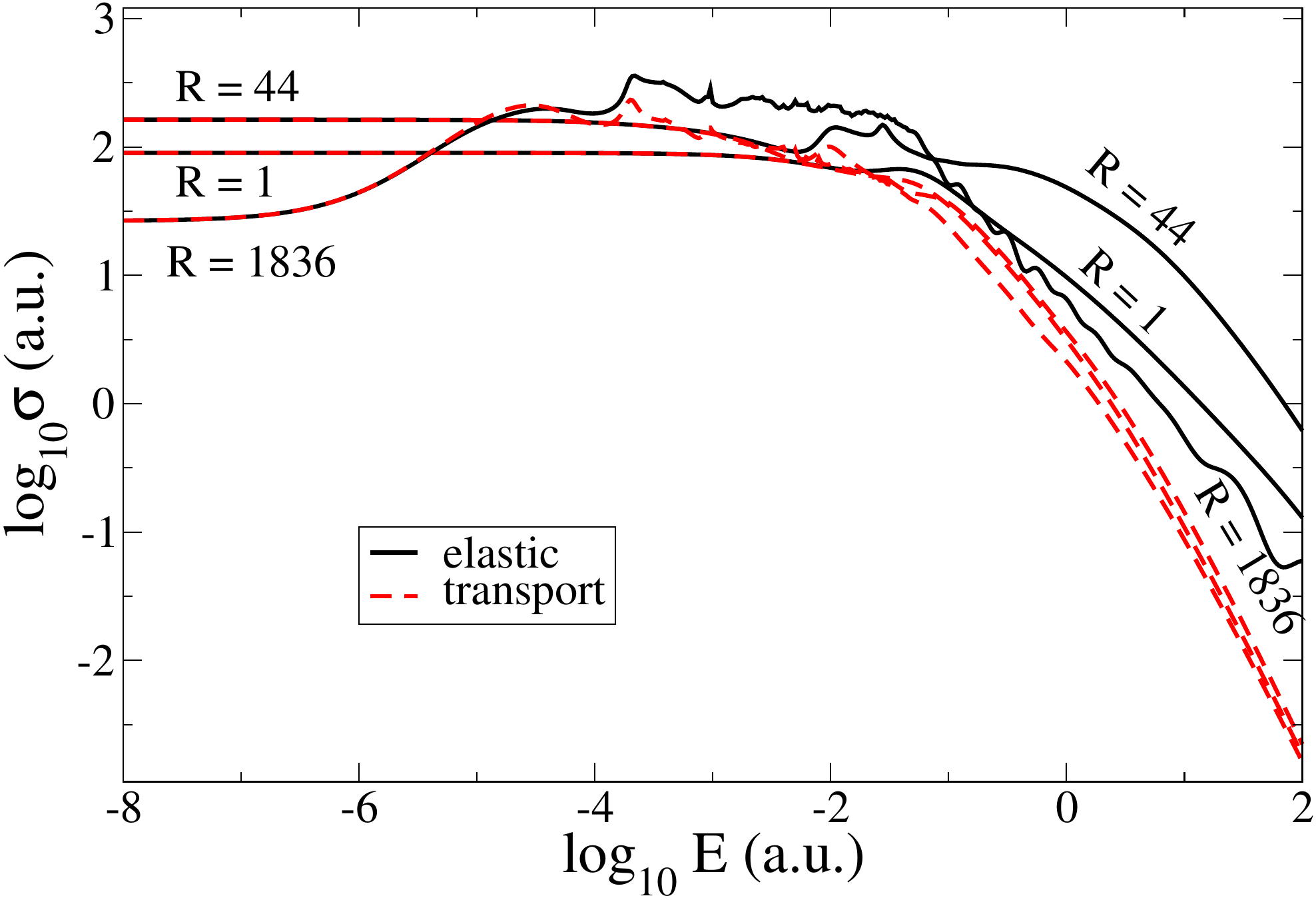}
\caption{Elastic (solid) and transport (dashed) dark atom
cross sections versus
energy for $R=1$, $44$ and $1836.15$
}
\label{fig:transport}
\end{figure}

To illustrate the difference between the elastic
and transport cross sections, we plot $\sigma_{\rm el}$ and
$\sigma'_t$ for a few representative values of $R$ in fig.\ 
\ref{fig:transport}.   As expected, at low energy where $s$-wave
scattering dominates, the two are equal, but they differ
at energies $E\gtrsim 0.1\,\epsilon_0$  where higher partial waves become important.  
We find that $\sigma'_t\propto 1/E$ for $0.1\,\epsilon_0 < E <
\epsilon_0$ and $\sigma'_t\propto 1/E^2$ at higher energies.  
(See the next section for a more detailed quantification of this
dependence.)  We use
this scaling in what follows in order to speed up computations of
$\sigma'_t$ above $0.1\,\epsilon_0$, since the addition of many
partial waves is time-consuming.  The asymptotic behavior
$\sigma'_t\propto 1/E^2$ is expected, since at high energies the scattering
is dominated by screened Coulomb scattering of the dark protons, which
shows exactly this energy dependence (up to logs).  We therefore
expect this scaling to be valid also for inelastic
contributions to the cross sections (such as from electronic
transitions) that are energetically allowed for $E > \epsilon_0$.
We disagree with the
assumption in \cite{CyrRacine:2012fz} that the cross sections drop
exponentially with $E$ for $E> \epsilon_0$, which contradicts our
expectations based on Coulomb scattering.

\begin{figure}[t]
\includegraphics[width=\columnwidth]{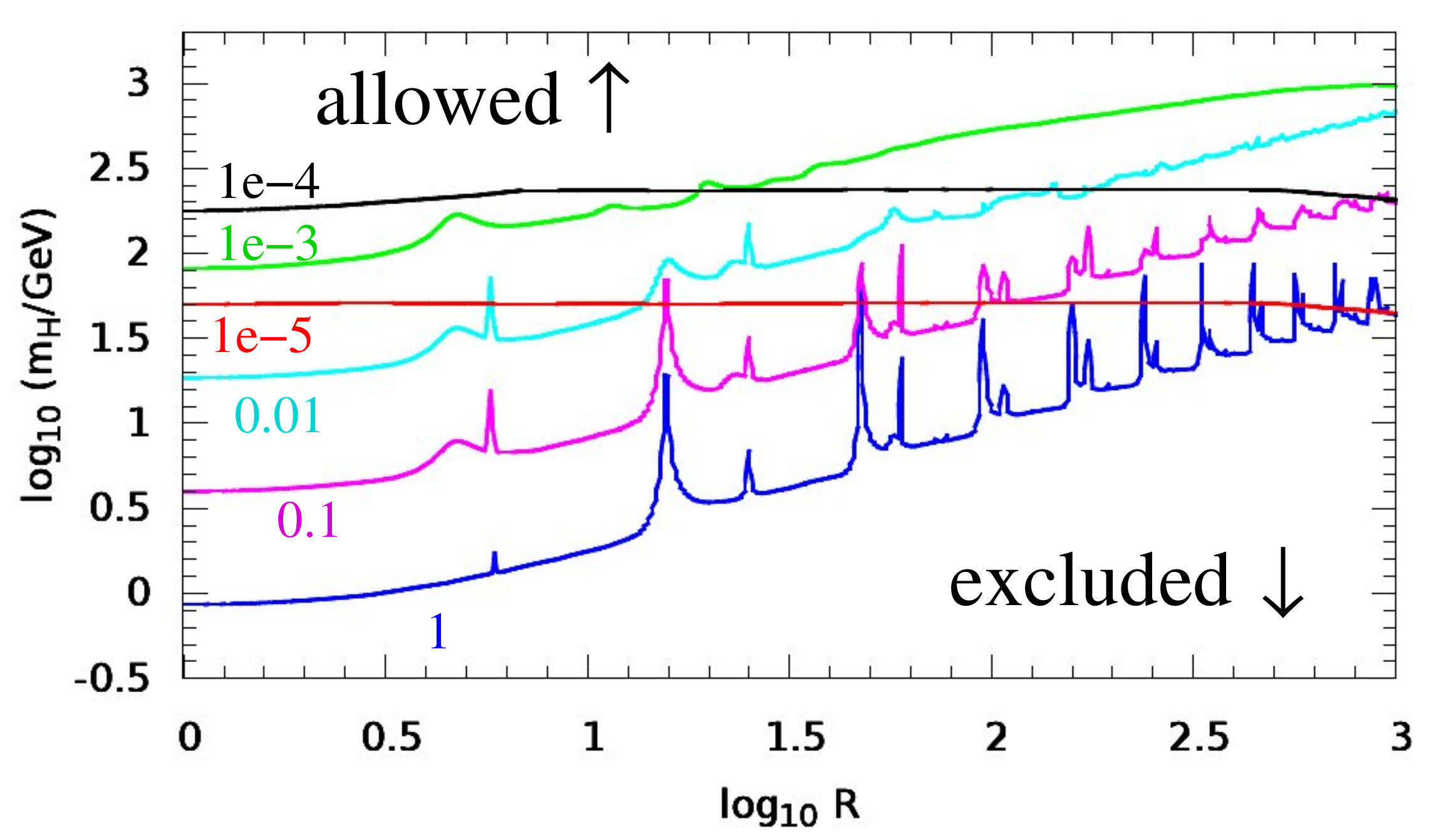}
\includegraphics[width=\columnwidth]{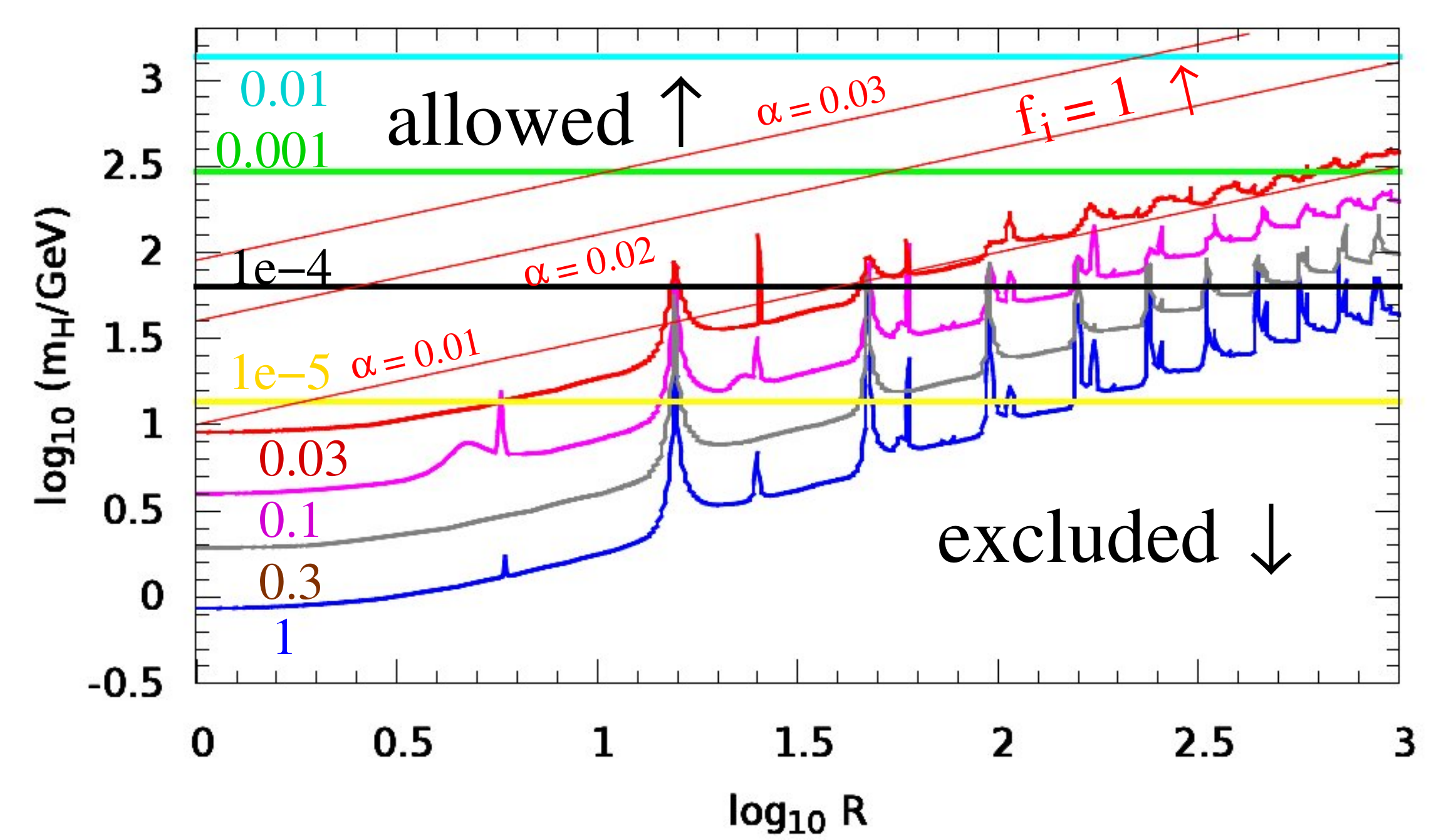}
\caption{(a) Uppermost: lower limit on the dark atom mass as a function of $R$,
from halo constraints on dark matter
self-interactions, assuming no ionization.
The curves are labeled by the value of $\alpha=1,\,0.1,\dots
10^{-5}$, which
is held fixed.\\  (b) Lower: modified limits, taking into account the
ionized fraction of dark atoms.  Thin diagonal lines indicate the boundary
above which the ionization fraction is $\sim 1$ for $\alpha = 0.01,\,
0.02,\,$ and $0.03$, according to eq.\ \ref{fieq} with $\xi=1$.
}
\label{fig:bounds}
\end{figure}

To constrain the parameter space of atomic DM, we 
impose the bound $\sigma'_t < 0.5\,$cm$^2/$g at 
several different DM velocities, 
$v=10,\,30,\,100,\,300,\,1000$ km/s, using $E = f(R)(v/c\alpha)^2\,
\epsilon_0$.
We scanned the $R$-$m_\dH$ plane for a range of $\alpha$ to find
upper limits on $m_\dH$ as a function of $R$.  
The results are shown in fig.\ \ref{fig:bounds}(a).   The constraints
show non-monotonic dependence on $\alpha$, which we can understand
as follows.  For very small $\alpha$ the binding energy is small so
the kinetic energy is large compared to the binding energy.  In this
regime the scatterings are essentially Coulomb scatterings between the
dark protons, and smaller $\alpha$ leads to less scattering.  But as
$\alpha$ is increased, the binding energy becomes larger than the
kinetic energy and the scatterings really involve the whole atoms.  Now
larger $\alpha$ means more tightly bound and therefore smaller atoms, 
hence a decreasing cross-section with increasing $\alpha$.
Alternatively, one could say that for small $\alpha$ the formation of
atoms fails to screen the Coulomb interaction, so scattering rates scale
as expected with coupling strength.  But as the coupling increases, the
charges are ever more effectively screened within atoms, and the
residual interactions get weaker with increasing $\alpha$.

In terms of dependence upon $R$,  we expect these results  
to be accurate for $R\gg 1$ where the Born-Oppenheimer
approximation holds.  Then the scattering potentials
are essentially independent of $R$ when expressed in atomic units, as
we have assumed.  This need no longer be the case when $R\sim 1$.

However we have unrealistically assumed up until now that there is
no significant fraction of dark ions.  In ref.\ \cite{Kaplan:2009de}
the ionization fraction $f_i$ was numerically determined over a range of
parameters.  We find that a good fit to their results is given by the
simple estimate
\be
	f_i \cong \rm{min}
	\left[10^{-10}\,\xi\,\alpha^{-4}\,R^{-1}\, (m_\dH/{\rm GeV})^2,\,
1\right]
\label{fieq}
\ee
where $\xi =  T_d/T_\gamma$, the present ratio of the
dark to visible photon temperatures, which was taken to be 1 in 
\cite{Kaplan:2009de}.  
This agrees with the result derived by ref.\ \cite{CyrRacine:2012fz},
which takes $\xi=0.4$,
while noting that the uncertainty in the estimate (\ref{fieq}) is
greater than the difference made by including the factor of $\xi$,
which we take to be 1 in the following.\footnote{In principle,
$\xi$ is a free parameter that is only determined by the relative
efficiency of reheating in the dark and visible sectors after
inflation, unless there are significant interactions between the
two sectors that we do not consider in this work.  Generically, one
would expect that $\xi\sim 1$ unless there is some (model-dependent)
reason for reheating only to the visible sector. 
Ref.\ \cite{CyrRacine:2012fz} show that big bang nucleosynthesis
bounds $\xi < 0.83-0.9$ at $3\sigma$, depending upon the number of
relativistic dark species at the time of BBN.}
It is then straightforward to show that the region of the
$R$-$m_\dH$ plane covered by fig.\ \ref{fig:bounds} corresponds to 
$f_i\sim 0$ for $\alpha>0.05$, while for $\alpha < 10^{-3}$, 
$f_i\sim 1$ over the entire region.  The boundaries above which $f_i$
becomes $\sim 1$ are shown as diagonal lines for the transition values
$\alpha= 0.01$, $0.02$, $0.03$ in fig.\ \ref{fig:bounds}(b).  Hence we
can ignore the effect of ionization on our constraints 
for $\alpha \gtrsim 0.03$, but it becomes
important at slightly lower values.  The transition is rather sudden due to
the high power of $1/\alpha$ in eq.\ (\ref{fieq}).

Ref.\ \cite{Feng:2009mn}  has considered the constraints from halo ellipticity and
from the Bullet Cluster on fully ionized atomic dark matter, 
numerically finding
the former to give a much stronger constraint, which we fit to the
form
\be
   {m_\dH\over {\rm GeV}} > \left(10^{6.7}\,\alpha\right)^{2/3}
\label{fengeq}
\ee	
 Rather
than computing a cross section and comparing it to a limiting value,
which is valid in the approximation that the scattering potential
can be modeled as a hard sphere (the assumption used in deriving the
limiting cross sections), ref.\ \cite{Feng:2009mn} compares the 
time needed to have several hard scatterings to the dynamical time
scale for the cluster as the criterion for erasure of ellipticity.
The Coulomb cross section is infrared divergent due to soft
scatterings, and even the momentum transfer cross section has a
logarithmic remnant of this divergence that gets cut off by the Debye
screening length of the DM plasma; hence the need for a specialized
treatment.  

The bounds we obtained for $\alpha < 0.1$ in 
fig.\ \ref{fig:bounds}(a) are thus superseded by (\ref{fengeq}),
indicated by the horizontal lines in the amended figure 
\ref{fig:bounds}(b), which also shows the elastic scattering bound for
additional values of $\alpha = 0.3$, $0.03$.  These bounds are quite
sensitive to the assumed value of the lowest velocity at which the
constraints are applicable, since they are determined by the region
of energies where $\sigma'_t \propto 1/E^2$.  They should thus be
considered as approximate, requiring a more detailed study of the 
effect of such a strongly velocity-dependent cross section on dwarf
galaxies, where $v\sim 10$ km/s applies.

\begin{figure}[t]
\includegraphics[width=\columnwidth]{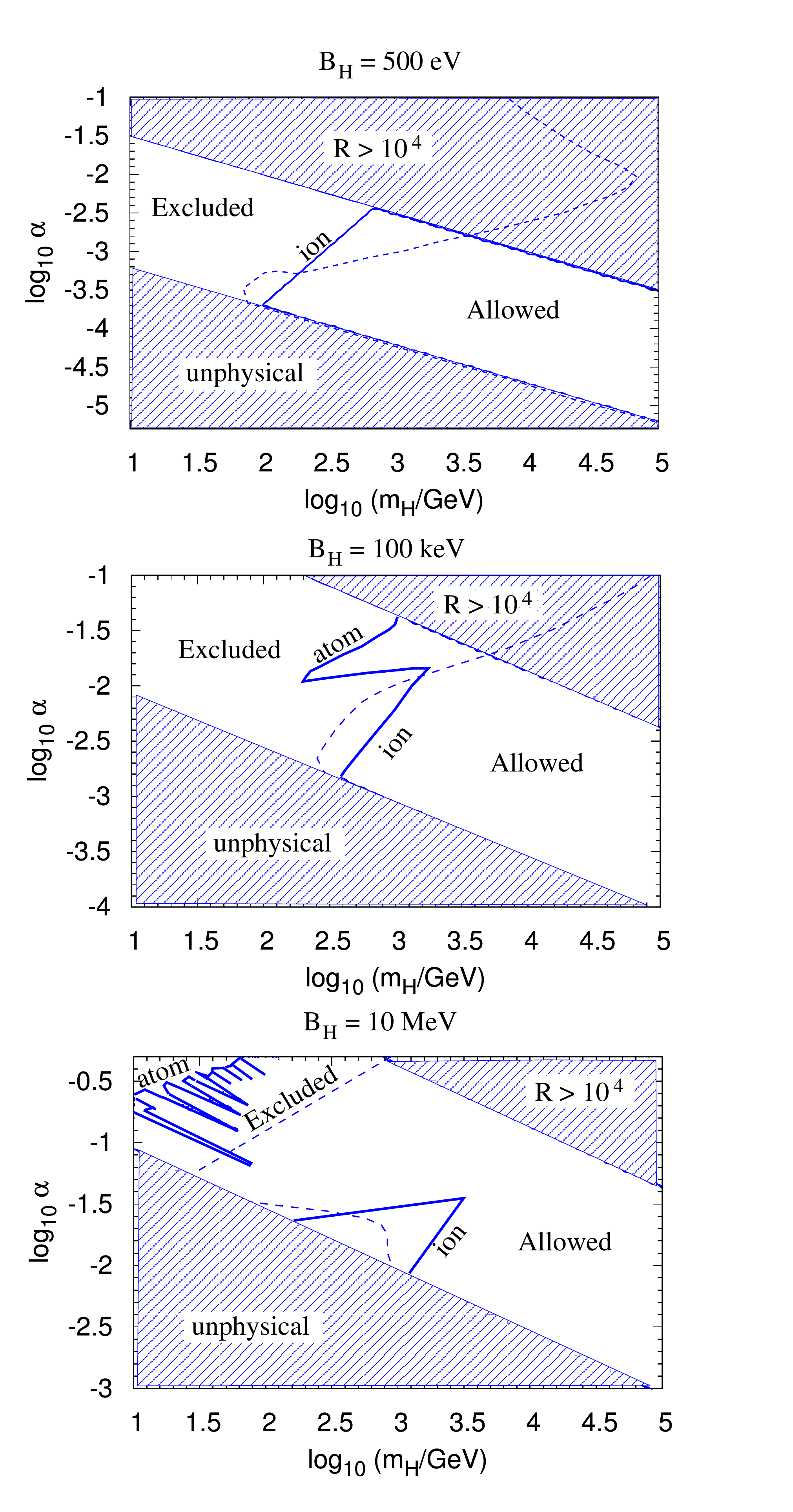}
\caption{Solid curves: boundary between allowed and excluded regions of atomic DM parameter space
for fixed values of the binding energy, $B_\dH = 500\,$ eV, 100 keV,
10 MeV, for comparison with fig.\ 20 of ref.\ \cite{CyrRacine:2012fz}, whose
constraints are given by the dashed curves.
}
\label{fig:sig-panel2}
\end{figure}

We also plot our constraints in the $m_\dH$-$\alpha$ plane for a few 
fixed values of the dark atom binding energy in fig.\
\ref{fig:sig-panel2}, for comparison with ref.\ \cite{CyrRacine:2012fz}
which presented their results in this way.  The parts of the constraints coming
from the fully ionized versus fully atomic forms are marked on the 
figure.  
We have computed the
ionization fraction using (\ref{fieq}), assuming $\xi=0.37$
as in \cite{CyrRacine:2012fz}.  Because it is numerically difficult
for us to compute the cross section for $R > 10^4$, we do not consider
these regions (so-labeled), in the upper right corners.  The lower
left corners are physically inaccessible since the binding energy is
given by $B_\dH = \frac12\alpha^2\,m_\dH / f(R)$ and $f(R)\equiv R+2+1/R$ 
cannot be less than $4$.  
As in fig.\ \ref{fig:bounds}(b),
we approximate the transition between ionized and atomic DM as sudden,
which explains the sharpness of the curves in the vicinity of $m_\dH = 
10^{3-3.5}$ in the lower two graphs of fig.\ \ref{fig:sig-panel2}.
(We expect the ion constraints to disappear for $f_i\lesssim 0.5$,
since in that case the halo ellipticity within the large atomic fraction
remains relatively undisturbed.)   In these
regions the constraint is coming from the fully ionized constituents.
On the other hand, the jaggedness of the constraint in the upper left
corner of the $B_\dH = 10$ MeV graph is a direct reflection of the 
strong $R$-dependence of the atomic scattering cross section, which 
was not taken into account in ref.\ \cite{CyrRacine:2012fz}.  We find that the allowed regions
are generally larger than given in that work.

\begin{table}[t]
 \begin{center}
\begin{tabular}{|r|c|c|c|c||r|c|c|c|c|}
\hline
$R$  & $a_0$ & $a_1$ & $a_2$ & $\chi^2$ & $R$  & $a_0$ & $a_1$ & $a_2$
& $\chi^2$\\
\hline 
1 & 0.011 & 0.221 & 0.063 & 0.084  & 400 & 0.005 & 0.290 & 0.049 &  1.000 \\
5 & 0.011 & 0.178 & 0.060 &  1.696 & 500 & 0.005 & 0.306 & 0.051 &  1.157 \\
 10 & 0.012 & 0.197 & 0.053 & 0.056 &  600 & 0.004 & 0.333 & 0.051 &  2.065 \\
 20 & 0.006 & 0.251 & 0.045 &  0.288 &  700 & 0.005 & 0.333 & 0.053 &  2.120 \\

 30 & 0.007 & 0.241 & 0.044 &  0.208 &  800 & 0.006 & 0.320 & 0.056 &  0.876 \\

 40 & 0.008 & 0.233 & 0.044 &  0.194 &   900 & 0.004 & 0.364 & 0.055 &  2.208 \\
 50 & 0.003 & 0.331 & 0.038 &  2.599 & 1000 & 0.007 & 0.318 & 0.060 & 0.593 \\
 60 & 0.005 & 0.277 & 0.041 &  1.026 & 1500 & 0.006 & 0.398 & 0.062 &  1.351 \\
 70 & 0.006 & 0.259 & 0.043 &  0.567 & 2000 & 0.008 & 0.407 & 0.069 & 2.964 \\
 80 & 0.006 & 0.251 & 0.043 &  0.412 & 2500 & 0.008 & 0.472 & 0.070 &  2.272 \\
 90 & 0.006 & 0.258 & 0.043 &  0.631 & 3000 & 0.003 & 0.697 & 0.062 & 4.531 \\
 100 & 0.003 & 0.325 & 0.039 & 2.726 & 3500 & 0.005 & 0.647 & 0.070 &  3.677 \\
 200 & 0.005 & 0.280 & 0.045  &  0.942 & 4000 & 0.002 & 0.970 & 0.059 & 10.014 \\
 300 & 0.005 & 0.281 & 0.047 &  0.972 & 4500 & 0.002 & 1.045 & 0.060 & 15.530 \\

\hline
\end{tabular}
\caption{Coefficients of the ansatz (\ref{fiteq}) that give the
best fit to the transport cross section for the given value of 
$R$.  The quality of the fit is indicated by $\chi^2$.
}
\label{tab:fits}
\end{center}
\end{table}

\begin{figure}[t]
\includegraphics[width=\columnwidth]{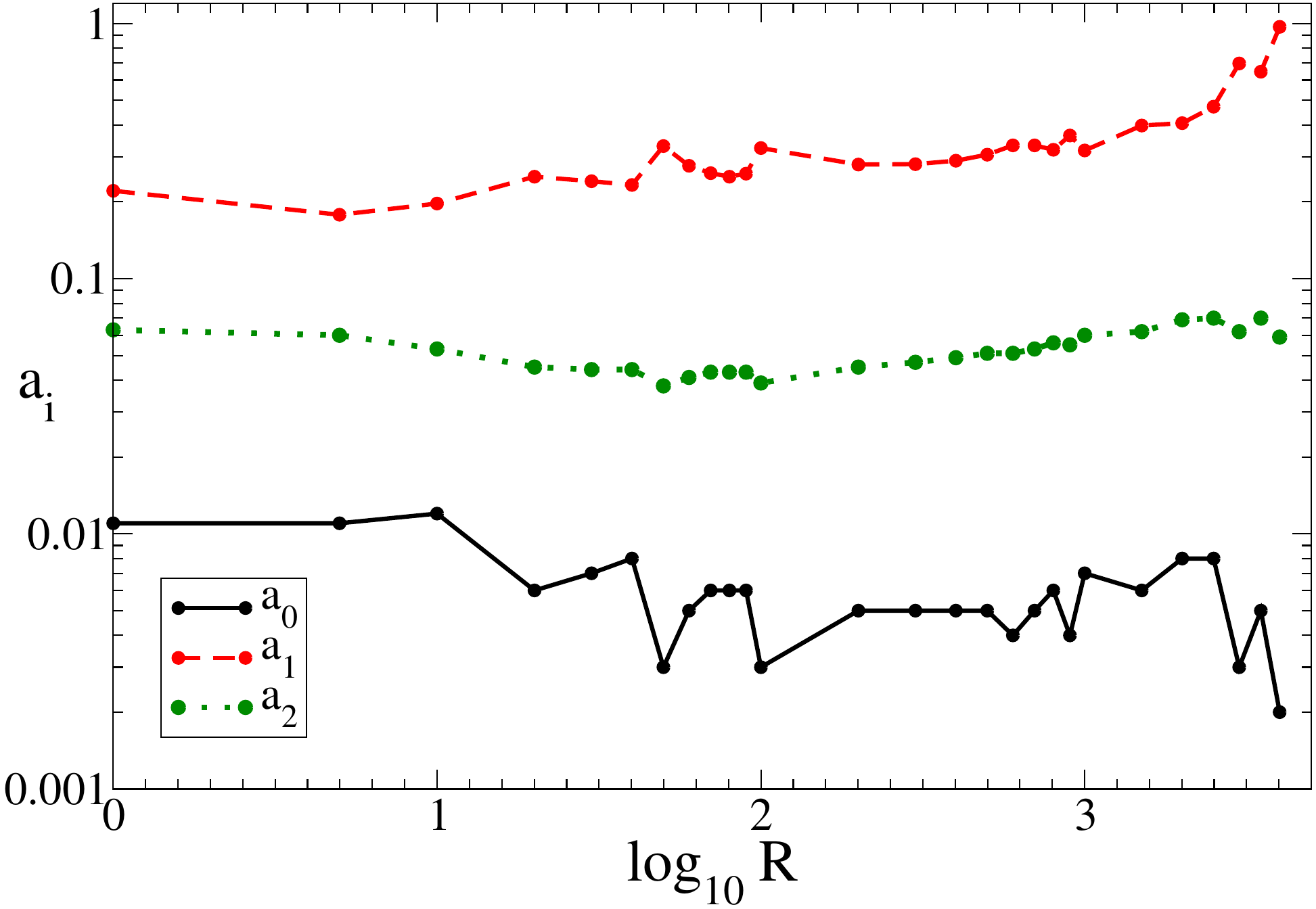}
\caption{Graphical representation of $R$-dependence of the
fit coefficients given in table \ref{tab:fits}.
}
\label{fig:coeff}
\end{figure}

\begin{figure*}[t]
\includegraphics[width=2\columnwidth]{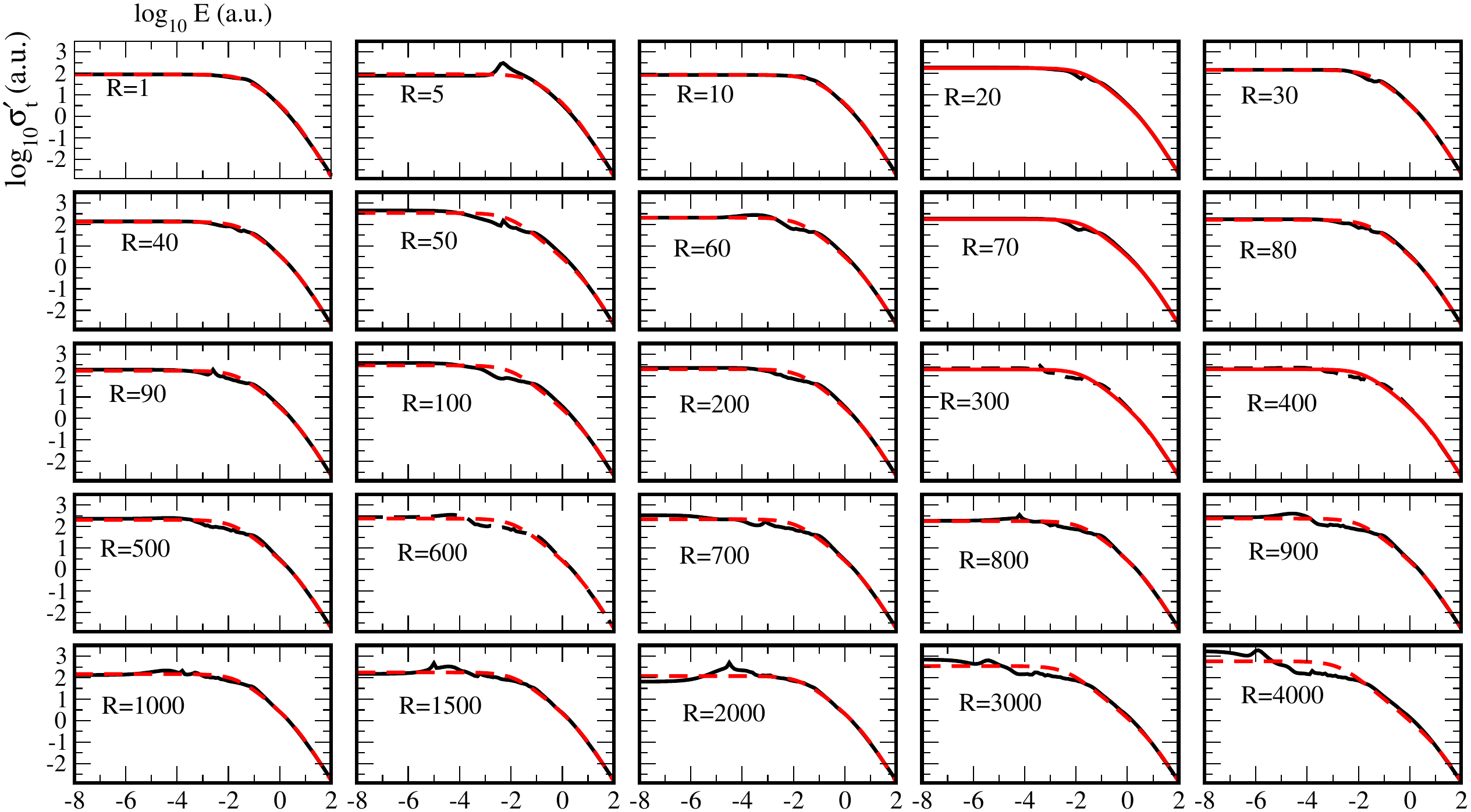}
\caption{Comparison of atomic momentum transfer cross section (solid)
lines with the best fit using ansatz (\ref{fiteq}) (dashed) for a 
range of $R$ values.  $\log_{10}\sigma_t'$ versus $\log_{10} E$ is 
plotted, in atomic units. 
}
\label{fig:fit-panel}
\end{figure*}

\section{Analytic fits to transport cross section}
\label{analytic}

As is apparent from fig.\ \ref{fig:panels}, it would be difficult to
give analytic formulas for the energy-dependence of the atomic
scattering cross sections in the cases where there are strong
resonances or vanishing scattering lengths.  On the other hand, 
there are many examples, such as the
cases $R=1$ and 44 shown in fig.\ \ref{fig:transport}, where 
$\sigma'_t$
has a rather simple dependence, which we find can be satisfactorily
fit by the ansatz
\be
	\sigma'_t \cong (a_0 + a_1 E + a_2 E^2)^{-1}
\label{fiteq}
\ee
where all quantities are expressed in atomic units.  We focus
here on the transfer cross section rather than the elastic one,
since it is the more physically relevant quantity for applications
such as those considered in the previous section.  
In table 
\ref{tab:fits} we give the best-fit values of the coefficients $a_i$
for the transport cross section, for a selection of $R$ values.  The
goodness of the fit is also given there as 
\be
	\chi^2 = \sum_{i=1}^{101} 
	\log_{10}^2(\sigma'_t/{\rm fit})
\ee
where the sum is over 101 uniformly spaced values of $\log_{10}(E)\in
[-8,2]$.   We plot the coefficients $a_i$ versus $R$ in fig.\ 
\ref{fig:coeff} to underscore that they tend to vary rather slowly
with $R$, especially $a_1$ which determines the behavior at
intermediate energies.  

The fits are graphically compared to the accurate cross sections in
fig.\ \ref{fig:fit-panel}.  On a log scale they all look rather
good, but the errors can be significant for examples with
$\chi^2\gtrsim 1$.  For example at $R=100$ with $\chi^2=1$, the
maximum error is a factor of 2 discrepancy at $E=10^{-2.5}$. As $R$
increases the accuracy tends to get worse.  At $R=4000$,
the fit is $4$ times greater than the actual cross section at $E =
10^{-4}$.  For $R=1$ on the other hand, the maximum error is only
$20\%$.  Unless one happens to choose a value of $R$ that  gives a
large resonance or a zero of the singlet channel scattering length,
a reasonable approximation to the transfer cross section can be
obtained by
interpolating  the above results.  For example at $R=15.5$ where
there is such a zero  (see fig.\ \ref{fig:panels}), we find
$\chi^2=130$ and the fit underestimates the actual $\sigma'_t$ by
two orders of magnitude at low energies (although it still does well
for $E\gtrsim 10^{-4}$).  In the case of a large resonance as in
$R=25.405$ we find $\chi^2=7.9$ with the error coming from 
energies at and below the resonance region, while the fit remains
good for $E > 10^{-5}$.

\section{Dark molecular $\dH_2$ abundance}
\label{molecules}

\def\HH{\mathrm{H}}

So far we have assumed that dark atoms do not predominantly combine to form the
analog of $\HH_2$ molecules.  In the cosmos, the proportion of real
$\HH_2$ molecules is small because the molecular binding energy $4.5$ eV is
less than the energy of Ly-$\alpha$ photons that were copiously
produced by young, massive, hot stars.  $\HH_2$ is thus easily
dissociated by a readily available form of radiation.  On the other
hand, it is slow to form because it has no electric dipole moment, and
the reaction $\HH+\HH\to$ $\HH_2+\gamma$ proceeds through an electric
quadrupole transition, occurring only once in every $10^5$ scatterings.
Much more efficient means of producing $\HH_2$ are the catalyzed
reactions 
($\HH+p\to \HH_2^+$, $\HH_2^+ +\HH \to \HH_2 +p$)
and 
($\HH+\HH^{-}\to\HH_2^-$, $\HH_2^-+\HH\to \HH_2
  + \HH^-$)
that rely upon a small ionized population.

In the dark universe, assuming no analog of weak interactions, there
will be no dark stars in the conventional sense that would produce
ionizing dark radiation.  Any stars that form from dark matter will 
be powered only by gravitational contraction as the protostellar cloud
slowly cools and pressure rises.  There will generically still be some
ionized fraction $f_i$ of dark atoms however, as given in eq.\
(\ref{fieq})  This
creates the potential for $\dH_2$ to become the prevalent form of
dark matter in such a scenario.  

If $R$ is large, the predomination of $\dH_2$ is undesirable,
because the rotational excitations of $\dH_2$ have
small energies, $E_r = \ell(\ell+1)/2I$ where $I\sim m_\dH a_0^2$
is the moment of inertia.  In atomic units, $E_r\sim \mu_\dH/m_\dH =
1/f(R)$.  Collisions of $\dH_2$ molecules with kinetic energies
greater than this can be inelastic, exciting the rotational states,
which can decay via quadrupole radiation.  The ensuing dissipation of
the DM kinetic energy will allow its halo to collapse in the same way
as luminous matter.  A complete study of this issue is beyond the
scope of this paper; for now we merely note that large values of $R$
might turn out to be untenable.

This leaves open the question of ``how large $R$ is really large?'' in
the  context of inelastic $\dH_2$ scattering.  Interestingly, we can
make a quantitative estimate using the machinery of the previous
sections.  
Because $H_2$ has no dipole moment, rotational or
ro-vibrational transitions involve electric quadrupole radiation, which
requires a bound state with $J\geq 2$.
Therefore, rotational and ro-vibrational emission is only possible if
there is at least one bound state in the $\ell=2$ channel.
By solving the Schr\"odinger equation 
(\ref{Schr}) at $E=0$ and $\ell=2$, we can identify the lowest value
of $R$ for which a $d$-wave bound state (indicated by a node in its wave
function) exists between two $\dH$ atoms.  It turns
out to be at $R=15.42$, close to the first pole of $a_s$.
This value of $R$ is large enough so that the Born-Oppenheimer
approximation is still reasonable;
hence we can expect it to be a fairly good estimate of the
value of $R$ below which no low-energy rotational transitions are 
available, and the ground state $\dH_2$ molecule is safe from making
dangerous inelastic transitions, even if it does dominate over dark 
atoms.

\begin{figure}[t]
\includegraphics[width=\columnwidth]{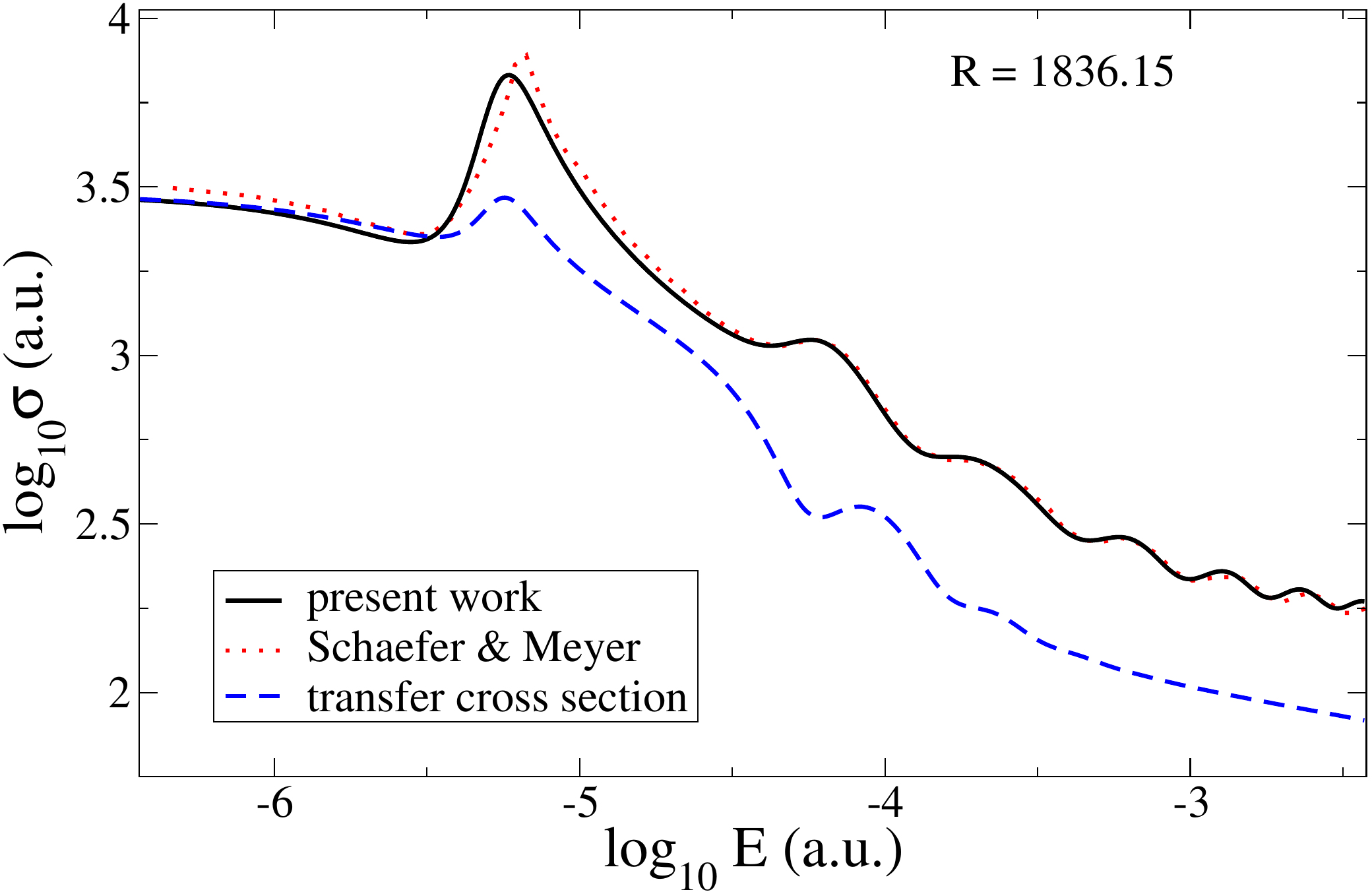}
\caption{Elastic scattering cross section of para-$\HH_2$ from
{\it ab initio} calculation of ref.\ \cite{Schaefer79} (dotted) and
our own calculation based upon the potential of 
\cite{Bauer76} (described in text).  Dashed curve shows  momentum
transfer cross section (present work).
}
\label{fig:mol-xsect-1836}
\end{figure}

\section{Scattering of dark $\dH_2$ molecules} 
\label{mole_xsect}

Given a potential
energy for $\dH_2$ self-interactions, we can use the same methodology
as for atoms to estimate the cross section for elastic $\dH_2$
scattering.  A number of {\it ab initio}  calculations of $\HH_2$-$\HH_2$
potentials have been given in the literature, as well as some
phenomenological ans\"atze that have been fit to physical properties
including the cross section.  At energies $E \gtrsim 1/R$, the
calculation is  complicated by the fact that the potential depends upon the
relative orientations of the two molecules.  At low energies where the
rotational states are not excited, one can use the spherically
symmetric term in the potential.   The Schr\"odinger equation for
molecular scattering differs from (\ref{Schr}) by the replacement
$f\to 2 f$ due to the mass of $\dH_2$.  Similarly the wave number is
given by $k = \sqrt{2fE}$ in atomic units.  The elastic cross section for para-$\HH_2$ scattering is given
by \cite{Bauer76} 
\be
	\sigma = {8\pi\over k^2}\sum_{\rm even\ \ell}(2\ell+1)
	\sin^2(\delta_\ell)
\label{mole-xsect}
\ee
with the extra factor of $2$ coming from the symmetry of the
scattering amplitude under $\theta\to\pi-\theta$ for identical
particles, as we also had for atomic scattering.

\begin{figure}[t]
\includegraphics[width=\columnwidth]{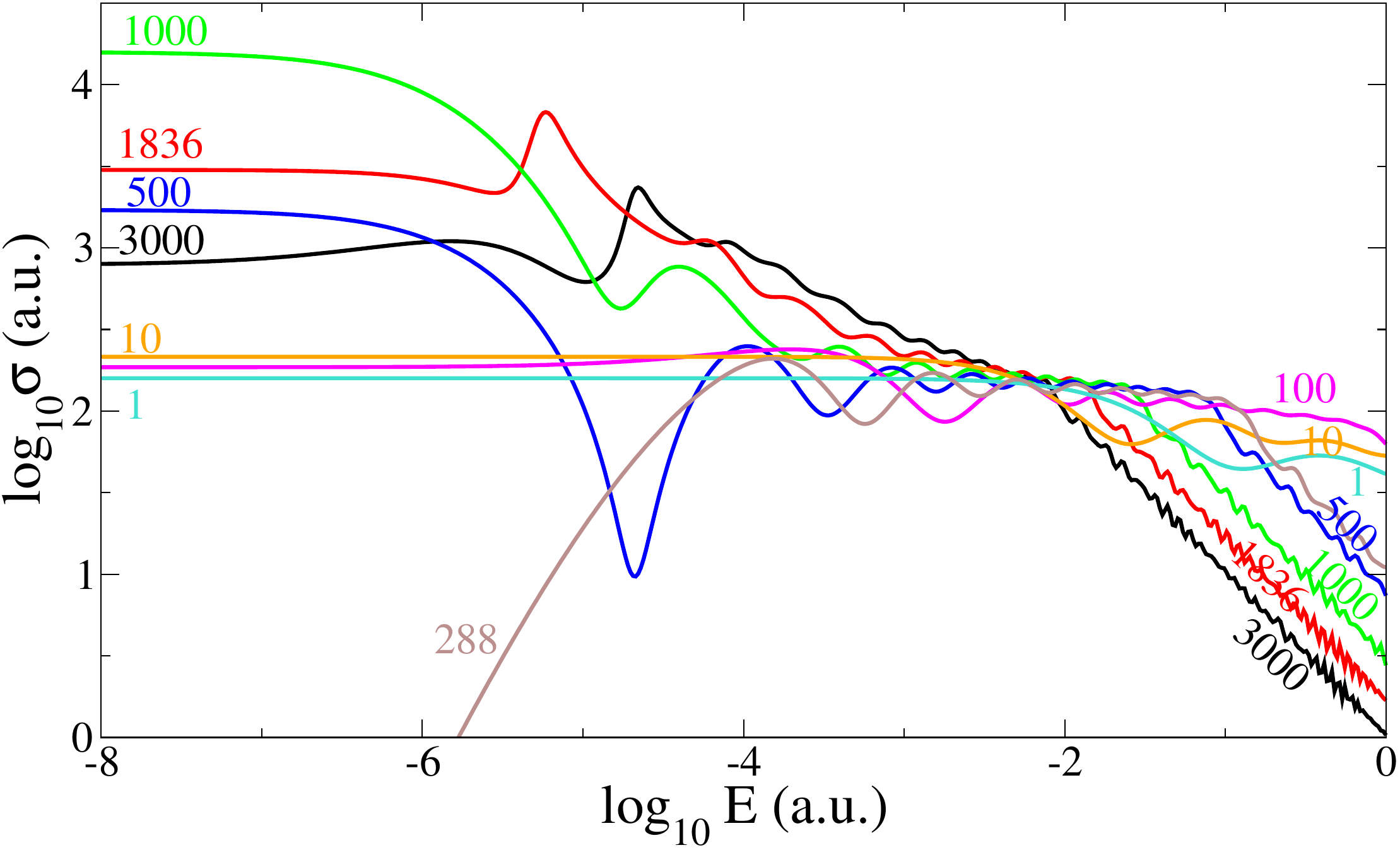}
\caption{Elastic cross sections for para-$\dH_2$ versus energy,
for $R=1$, 10, 100, 288, 500, 1000, 1836 and 3000.
}
\label{fig:mole-scatt}
\end{figure}

\begin{figure}[t]
\includegraphics[width=0.9\columnwidth]{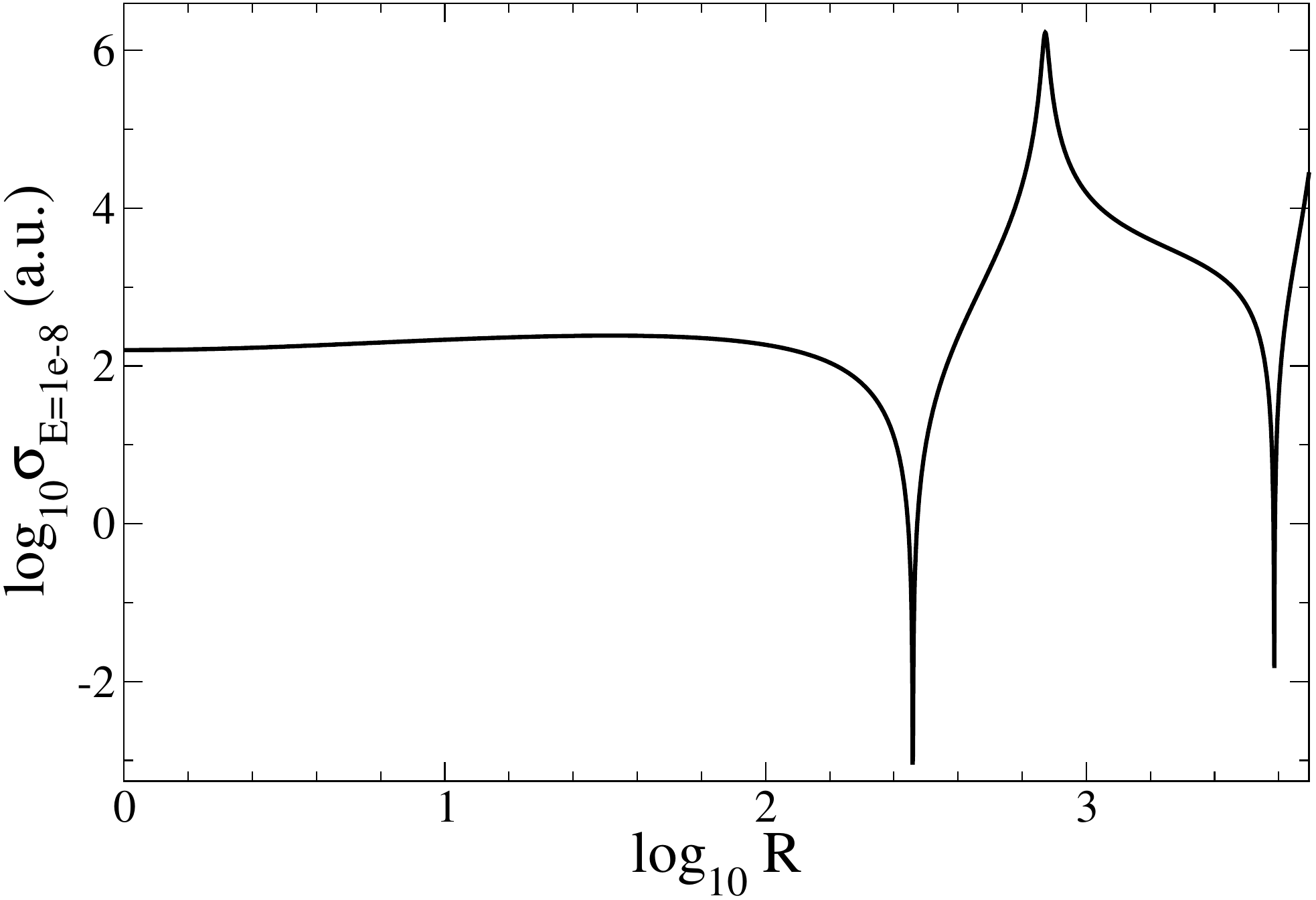}
\caption{Elastic cross section for para-$\dH_2$ versus $R$,
at energy $E=10^{-8}\epsilon_0$.
}
\label{fig:mol_x_sect-v-R}
\end{figure}

It is possible to obtain
a good description of experimentally measured cross sections for
$\HH_2$-$\HH_2$ scattering 
at low energies with
a potential of the same form as (\ref{silv}), using for
para-$\HH_2$ the parameter
values $c_0 = 3.778$, $c_1 = 1.947$, $c_2 = 3.763\times 10^{-3}$,
$C_6 = 12.0$, $C_8=239.9$, $C_{10} = 0$ \cite{Bauer76}.   Ref.\ 
\cite{Bauer76} does not include the $D$ factor, needed to keep the
long-distance part of the potential from contributing as $r\to 0$,
but we find that using $r_1 = 4$ gives satisfactory suppression
without changing the behavior near the shallow
minimum of the potential, at $r_m\sim 6.5\,a_0$ with $V_m\cong
-10^{-4}\,\epsilon_0$.  We plot the resulting cross section in fig.\
\ref{fig:mol-xsect-1836}, along with the result of ref.\ 
\cite{Schaefer79} based upon an {\it ab initio} determination of the
orientationally averaged potential.  The results are in fair
agreement, with a 10\% discrepancy at low $E$ which is due to the difference between the large-$r$ part of the Bauer {\it et al.}
potential \cite{Bauer76} we have adopted, and that assumed in ref.\ 
\cite{Schaefer79}.

\begin{figure}[t]
\includegraphics[width=\columnwidth]{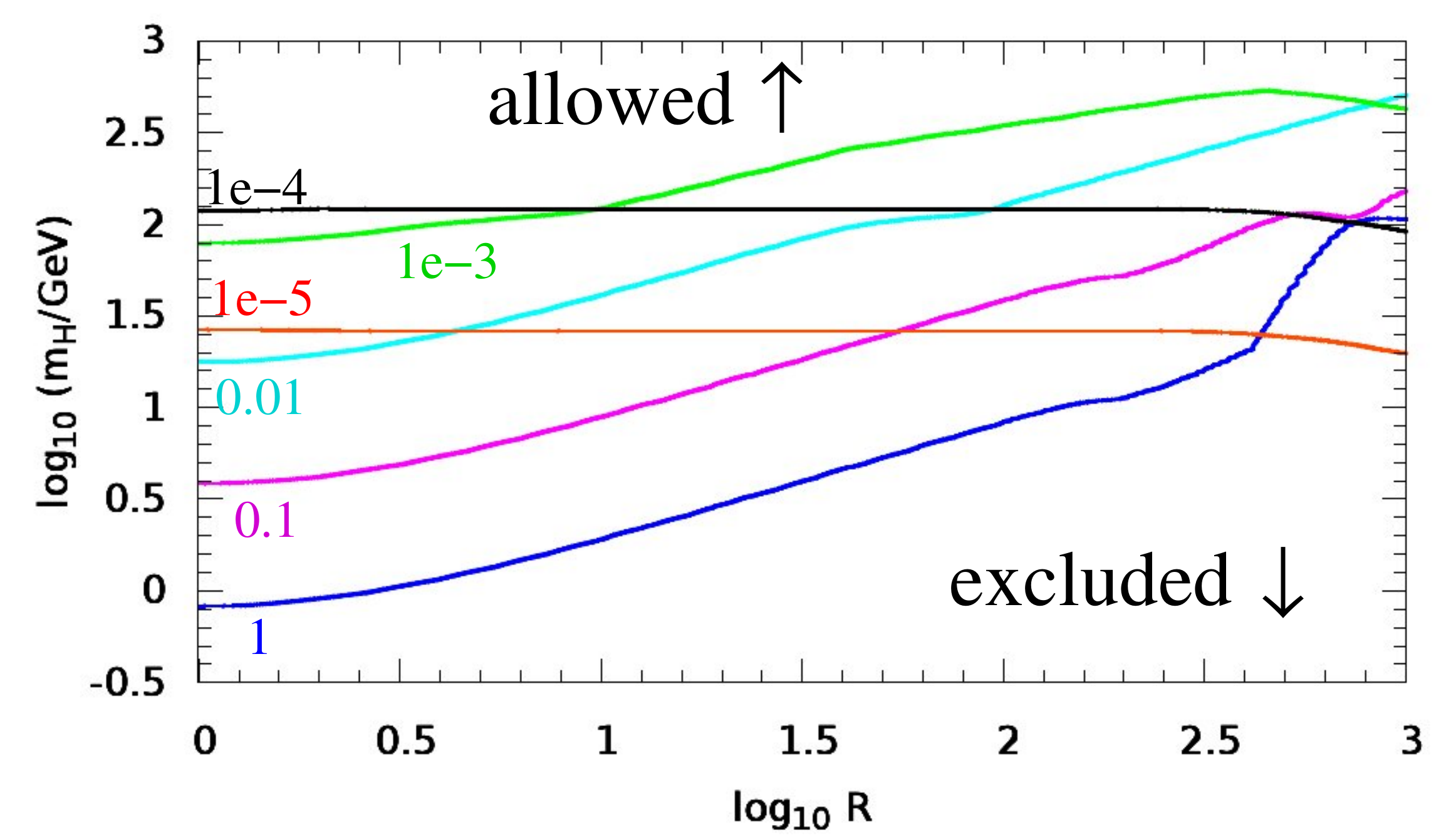}
\caption{Upper limits on $m_\dH$ versus $R$ from halo structure
for molecular dark matter, at $\alpha=1,\,0.1,\dots,10^{-5}$.
The mass plotted is still that of the atom, $m_\dH$.
}
\label{fig:bound-mole-el}
\end{figure}

Having reproduced known results at $R=1836$, we explore the dependence
of $\sigma(E)$ on $R$ for dark molecules.  A sample of cross sections
for representative values of $R$ is given in fig.\ \ref{fig:mole-scatt}.
Like for atoms, the cross sections generically approach a constant at
low energies, from the $s$-wave contribution, and start to exhibit
structure from the higher partial waves at energies $\epsilon_2\sim
R^{-3/2}\epsilon_0$.  An exceptional case is shown for $R=288$, close to the first zero
of the scattering length, in which the constant behavior is delayed
until smaller energies.  
A complementary view is given
in fig.\ \ref{fig:mol_x_sect-v-R}, which plots $\sigma$ at
$E=10^{-8}\epsilon_0$ as a function of $R$.  Because the potential is
quite shallow, there is only one bound state (giving a divergence of the
scattering length when its energy vanishes) for $R < 4000$.  The
weakly bound state enhances the cross section for $R\sim 1000$, making
it an order of magnitude or more larger than the typical value for 
dark atoms in this region of $R$.

Following the same procedure as for atomic dark matter, we have
estimated the constraints from structure formation
on the dark atom mass in the case where it is in molecular form.
The momentum transfer cross section is given again by replacing
$(2\ell+1)\to (3/2)(\ell+1)(\ell+2)/(2\ell+3)$ and $\delta_\ell\to
\delta_\ell-\delta_{\ell+2}$  in eq.\ 
(\ref{mole-xsect}), the effect of which is indicated in
fig.\ \ref{fig:mol-xsect-1836}.
The resulting bounds, shown in fig.\ \ref{fig:bound-mole-el}, are 
rather
similar to those we found for dark atoms in fig.\ \ref{fig:bounds}(a),
except for the absence of sharp features, thanks to the relative
smoothness of the molecular scattering length as a function of $R$
(see fig.\ \ref{fig:mol_x_sect-v-R}).  The bounds for molecular
dark matter are stronger at $R\sim 1000$ and $\alpha\sim 1$ 
than for atomic DM because
of the larger cross section at low energies.
Like in the case of dark atoms, we expect the constraints for
$\alpha \lesssim 10^{-2}$ to be stronger than shown here, since
the ionization fraction is estimated to be large and the assumption 
of domination by the molecular state will not be correct. 
Nevertheless we show them for comparison with fig.\
\ref{fig:bounds}(a).  The constraints from the ionized fraction at
small $\alpha$ will be the same as in fig.\
\ref{fig:bounds}(b).

In deriving these constraints, we have neglected the inelastic
contributions from ro-vibrational transitions that become
energetically allowed for $E\gtrsim R^{-1}$.  Ref.\ \cite{Quemener09}
shows that these  are individually much smaller than the
elastic cross section.  For example excitations from the ground state
to the lowest rotational states have cross sections of $\sim 9\,$a.u.\
at $E = 0.04\,\epsilon_0$, while transitions to the next lowest
excitations have cross sections an order of magnitude smaller.  At
this energy, the elastic cross section is still 40$\,$a.u.  Thus the
elastic part may be a better estimate of the total cross
section than one might have guessed.

\section{Summary and conclusions}
\label{conclusions}

We have computed the cross sections for elastic scattering of 
dark atoms and molecules, whose properties are analogous to those
of the visible world, and determined by the coupling strength
$\alpha$, the atom mass $m_\dH$, and the ratio $R$ of dark proton and
electron masses.  In a world with $R=1836.15$, and assuming $\alpha\ll
1$, there would be nothing to do, since the properties of dark atoms
and molecules would be identical to those of their visible
counterparts once expressed in the atomic units of length
$a_0=(\alpha\mu)^{-1}$ and energy $\epsilon_0 = \alpha^2\mu$.  The
nontrivial part of our job was to investigate how scattering
changes as a function of $R$.  Fortunately, for $R\gg 1$, the
Born-Oppenheimer approximation tells us that the interaction
potentials (in atomic units) do not depend upon $R$.
All the $R$-dependence is kinematic and appears in the Schr\"odinger
equation.  By solving the Schr\"odinger equation using
accurate determinations
for the potentials, we are able to make quantitative predictions
for dark atom scattering at $R\gg 1$.

We found that the cross sections for atom-atom scattering depend very
strongly upon $R$, due to the number of bound states of the singlet
scattering channel changing rapidly with $R$, with consequent
divergences (and zeroes) in the singlet channel scattering length.
The triplet channel has a shallower potential and thus less pronounced
$R$-dependence.  The same is true for scattering of dark molecules,
whose interaction potential is also shallow.  Our exploration of the
cosmology of dark molecules, though cursory, is the first one in the
literature that we are aware of, and may lay useful groundwork for
further study.  One conclusion is that dark molecules may be
disfavored for $R \gtrsim 15$ since in that case the inelastic
scattering into rotationally excited states could make the DM too
dissipative to remain in an extended halo.

As an application, we determined constraints from self-interactions on the atomic dark matter
parameter space following from observations of halo ellipticity and
central densities of dwarf spheroidal galaxies.  Moreover we have
given simple analytic fits to the energy dependence of the
momentum-transfer cross sections
that are accurate to $20\%$ in some cases (despite the general
complexity of the functions being modeled), and good enough for
order of magnitude estimates in many other cases. 
These results improve
upon previous ones in the literature by virtue of our more accurate cross
sections, with respect to energy- and $R$-dependence, and by properly
distinguishing between the elastic and momentum transfer cross
sections.  In addition to constraints, there are suggestions that such
self-interactions could be useful for addressing discrepancies between
predictions of cold dark matter and some aspects of observed structure
formation on small scales.  One could thus anticipate that some of the
borderline regions may actually be favored.  We will address this
issue in more detail in an upcoming paper.

\bigskip
{\bf Acknowledgments.}  We thank Francis-Yan Cyr-Racine, Gil Holder,
and Kris Sigurdson for helpful discussions or correspondence.
JC thanks the Aspen Center for Physics for its congenial working
environment while this research was in progress.


\begin{thebibliography}{10}


\bibitem{Hodges:1993yb} 
  H.~M.~Hodges,
  Phys.\ Rev.\ D {\bf 47}, 456 (1993).

\bibitem{Goldberg:1986nk} 
  H.~Goldberg and L.~J.~Hall,
  Phys.\ Lett.\ B {\bf 174}, 151 (1986).

\bibitem{Berezhiani:1995yi} 
  Z.~G.~Berezhiani and R.~N.~Mohapatra,
  Phys.\ Rev.\ D {\bf 52}, 6607 (1995)
  [hep-ph/9505385].

\bibitem{Berezhiani:1995am} 
  Z.~G.~Berezhiani, A.~D.~Dolgov and R.~N.~Mohapatra,
  Phys.\ Lett.\ B {\bf 375}, 26 (1996)
  [hep-ph/9511221].


\bibitem{Foot:1995pa} 
  R.~Foot and R.~R.~Volkas,
  Phys.\ Rev.\ D {\bf 52}, 6595 (1995)
  [hep-ph/9505359].


\bibitem{Mohapatra:2000qx} 
  R.~N.~Mohapatra and V.~L.~Teplitz,
  Phys.\ Rev.\ D {\bf 62}, 063506 (2000)
  [astro-ph/0001362].

\bibitem{Foot:2004pa} 
  R.~Foot,
  Int.\ J.\ Mod.\ Phys.\ D {\bf 13}, 2161 (2004)
  [astro-ph/0407623].


\bibitem{Strassler:2006im} 
  M.~J.~Strassler and K.~M.~Zurek,
  Phys.\ Lett.\ B {\bf 651}, 374 (2007)
  [hep-ph/0604261].

\bibitem{ArkaniHamed:2008qn} 
  N.~Arkani-Hamed, D.~P.~Finkbeiner, T.~R.~Slatyer and N.~Weiner,
  Phys.\ Rev.\ D {\bf 79}, 015014 (2009)
  [arXiv:0810.0713 [hep-ph]].


\bibitem{Kaplan:2009de} 
  D.~E.~Kaplan, G.~Z.~Krnjaic, K.~R.~Rehermann and C.~M.~Wells,
  JCAP {\bf 1005}, 021 (2010)
  [arXiv:0909.0753 [hep-ph]].

\bibitem{Behbahani:2010xa} 
  S.~R.~Behbahani, M.~Jankowiak, T.~Rube and J.~G.~Wacker,
  Adv.\ High Energy Phys.\  {\bf 2011}, 709492 (2011)
  [arXiv:1009.3523 [hep-ph]].

\bibitem{Kaplan:2011yj} 
  D.~E.~Kaplan, G.~Z.~Krnjaic, K.~R.~Rehermann and C.~M.~Wells,
  JCAP {\bf 1110}, 011 (2011)
  [arXiv:1105.2073 [hep-ph]].

\bibitem{Cline:2012is} 
  J.~M.~Cline, Z.~Liu and W.~Xue,
  Phys.\ Rev.\ D {\bf 85}, 101302 (2012)
  [arXiv:1201.4858 [hep-ph]].

\bibitem{Cline:2012ei} 
  J.~M.~Cline, Z.~Liu and W.~Xue,
  Phys.\ Rev.\ D {\bf 87}, 015001 (2013)
  [arXiv:1207.3039 [hep-ph]].

\bibitem{CyrRacine:2012fz} 
  F.~-Y.~Cyr-Racine and K.~Sigurdson,
  Phys.\ Rev.\ D {\bf 87}, 103515 (2013)
  [arXiv:1209.5752 [astro-ph.CO]].

\bibitem{Cyr-Racine:2013fsa} 
  F.~-Y.~Cyr-Racine, R.~de Putter, A.~Raccanelli and K.~Sigurdson,
  arXiv:1310.3278 [astro-ph.CO].

\bibitem{Weinberg:2013aya} 
  D.~H.~Weinberg, J.~S.~Bullock, F.~Governato, R.~K.~de Naray and A.~H.~G.~Peter,
  arXiv:1306.0913 [astro-ph.CO].

\bibitem{KW74}
W.\ Kolos, L.\ Wolniewicz, Chem.\ Phys.\ Lett.\ {\bf 24}, 457 (1974)

\bibitem{Silvera}
I.F.\ Silvera, Rev.\ Mod.\ Phys.\ {\bf 52}, 393 (1980);\\
I.F.\ Silvera, J.M.\ Walraven, Prog.\ Low Temp.\ Phys.\ {\bf 10}, 139
(1986)

\bibitem{Wol93}
L.\ Wolniewicz, J.\ Chem.\ Phys.\ {\bf 99}, 1851 (1993)

\bibitem{Joudeh}
B.R.\ Joudeh, Physica {\bf B421}, 41 (2013)

\bibitem{Jam99}
M.J.\ Jamieson {\it et al.}, Phys.\ Rev.\ {\bf A61}, 014701 (1999)

\bibitem{Chakra06}
A.\ Sen, S.\ Chakraborty, A.S.\ Ghosh, Europhys.\ Lett.\ {\bf 76},
582 (2006)

\bibitem{Will95}
C.J.\ Williams, P.S.\ Julienne, Phys.\ Rev.\ {\bf A47}, 1524 (1995)

\bibitem{Jam98} M.J.\ Jamieson and A.\ Dalgarno, J.\ Phys.\ B:
At.\ Mol.\ Opt.\ Phys.\ {\bf 31} L219 (1998)

\bibitem{Chakra07} S.\ Chakraborty, A.\ Sen and A.S.\ Ghosh,
Eur.\ Phys.\ J.\ D {\bf 45} 261 (2007)

\bibitem{Fox67} J.W.\ Fox and E.\ Gal, Proc.\ Phys.\ Soc.\ {\bf 90}
55 (1967)

\bibitem{Krstic99}
P.\ Krstic and D.\ Schultz, J.\ Phys.\ B: At.\ Mol.\ Opt.\ 
Phys.\ {\bf 32} (1999) 3485

\bibitem{MiraldaEscude:2000qt} 
  J.~Miralda-Escude,
  Ap.~J.~{\bf 564} (2002) 60
  [astro-ph/ 0002050].

\bibitem{Peter:2012jh} 
  A.~H.~G.~Peter, M.~Rocha, J.~S.~Bullock and M.~Kaplinghat,
  arXiv:1208.3026 [astro-ph.CO].

\bibitem{Markevitch:2003at} 
  M.~Markevitch, A.~H.~Gonzalez, D.~Clowe, A.~Vikhlinin, L.~David, W.~Forman, C.~Jones and S.~Murray {\it et al.},
  Astrophys.\ J.\  {\bf 606}, 819 (2004)
  [astro-ph/0309303].

\bibitem{Randall:2007ph} 
  S.~W.~Randall, M.~Markevitch, D.~Clowe, A.~H.~Gonzalez and M.~Bradac,
  Astrophys.\ J.\  {\bf 679}, 1173 (2008)
  [arXiv:0704.0261 [astro-ph]].

\bibitem{Ostriker:1999ee} 
  J.~P.~Ostriker,
  Phys.\ Rev.\ Lett.\  {\bf 84}, 5258 (2000)
  [astro-ph/9912548].

\bibitem{Hennawi:2001be} 
  J.~F.~Hennawi and J.~P.~Ostriker,
  [astro-ph/0108203].

\bibitem{Gnedin:2000ea} 
  O.~Y.~Gnedin and J.~P.~Ostriker,
  [astro-ph/0010436].

\bibitem{SanchezSalcedo:2005vc} 
  F.~J.~Sanchez-Salcedo,
  Astrophys.\ J.\  {\bf 631}, 244 (2005)
  [astro-ph/0506345].

\bibitem{Zavala:2012us} 
  J.~Zavala, M.~Vogelsberger and M.~G.~Walker,
  arXiv:1211.6426 [astro-ph.CO].

\bibitem{Tulin:2013teo} 
  S.~Tulin, H.~-B.~Yu and K.~M.~Zurek,
  Phys.\ Rev.\ D {\bf 87}, 115007 (2013)
  [arXiv:1302.3898 [hep-ph]].

\bibitem{PDR03}
E.R.\ Cohen, D.\ Lide, G.\ Trigg, ``Physicist's Desk Reference,''
Springer (2003)

\bibitem{Flann74}
M.R.\ Flannery and K.J.\ McCann, Phys.\ Rev.\ {\bf A9}, 1947 (1974)

\bibitem{Feng:2009mn} 
  J.~L.~Feng, M.~Kaplinghat, H.~Tu and H.~-B.~Yu,
  JCAP {\bf 0907}, 004 (2009)
  [arXiv:0905.3039 [hep-ph]].

\bibitem{Bauer76}
W.\ Bauer, B.\ Lantzsch1, J.P.\ Toennies and  K.\ Walaschewski,
Chem.\ Phys.\ {\bf 17}, 19 (1976)

\bibitem{Schaefer79}
J.\ Schaefer and W.\ Meyer, J.\ Chem.\ Phys.\ {\bf 70}, 344 (1979)


\bibitem{Quemener09}
G.\ Qu\'em\'ener and N.\ Balakrishnan, J.\ Chem.\ Phys.\ {\bf 130},
114303 (2009)

\end{thebibliography}
\end{document}